\documentclass[10pt,letterpaper]{article}
\usepackage[top=0.85in,left=2.75in,footskip=0.75in]{geometry}

\usepackage{amsmath,amssymb}

\usepackage{changepage}

\usepackage{textcomp,marvosym}

\usepackage{cite}

\usepackage{nameref,hyperref}

\usepackage[right]{lineno}

\usepackage[nopatch=eqnum]{microtype}
\DisableLigatures[f]{encoding = *, family = * }

\usepackage[table]{xcolor}

\usepackage{array}

\newcolumntype{+}{!{\vrule width 2pt}}

\newlength\savedwidth

\raggedright
\usepackage[aboveskip=1pt,labelfont=bf,labelsep=period,justification=raggedright,singlelinecheck=off]{caption}

\makeatletter
\renewcommand{\@biblabel}[1]{\quad#1.}
\makeatother

\usepackage{bm} 
\usepackage{subcaption} 
\usepackage{booktabs}
\usepackage{xltabular}
\usepackage{longtable} 
\usepackage[title]{appendix}
\usepackage{placeins} 
\usepackage{rotating} 

\newcommand{\CellWithForcedBreak}[2][c]{
\begin{tabular}[#1]{@{}c@{}}#2\end{tabular}}

\usepackage{lastpage,fancyhdr,graphicx}
\usepackage{epstopdf}
\fancyheadoffset[L]{2.25in}
\fancyfootoffset[L]{2.25in}
\begin{document}
\vspace*{0.2in}

\begin{flushleft}
{\Large
\textbf{Dynamic modelling and evaluation of preclinical trials in acute leukaemia} 
}
\newline
\\
Julian W\"asche\textsuperscript{1},
Romina Ludwig\textsuperscript{2,3},
Irmela Jeremias\textsuperscript{2,3,4},
Christiane Fuchs\textsuperscript{1,5*}
\\
\bigskip
\textbf{1} Data Science Group, Faculty of Business Administration and Economics, Bielefeld University, Bielefeld, North Rhine-Westphalia, Germany
\\
\textbf{2} Research Unit Apoptosis in Hematopoietic Stem Cells, Helmholtz Munich, German Research Center for Environmental Health (HMGU), Munich, Bavaria, Germany
\\
\textbf{3} Department of Pediatrics, Dr. von Hauner Children’s Hospital, University Hospital, LMU Munich, Munich, Bavaria, Germany
\\
\textbf{4} German Cancer Consortium (DKTK), partner site Munich, a partnership between DKFZ and University Hospital LMU Munich, Munich, Bavaria, Germany
\\
\textbf{5} Institute of Computational Biology, Computational Health Center, Helmholtz Munich, Neuherberg, Bavaria, Germany
\\
\bigskip

* christiane.fuchs@uni-bielefeld.de

\end{flushleft}

\section*{Abstract}
Dynamic models are widely used to mathematically describe biological phenomena that evolve over time. One important area of application is leukaemia research, where leukaemia cells are genetically modified in preclinical studies to explore new therapeutic targets for reducing leukaemic burden. In advanced experiments, these studies are often conducted in mice and generate time-resolved data, the analysis of which may reveal growth-inhibiting effects of the investigated gene modifications. However, the experimental data is oftentimes evaluated using statistical tests which compare measurements from only two different time points. This approach does not only reduce the time series to two instances but also neglects biological knowledge about cell mechanisms. Such knowledge, translated into mathematical models, expands the power to investigate and understand effects of modifications on underlying mechanisms based on experimental data.
We utilise two population growth models -- an exponential and a logistic growth model -- to capture cell dynamics over the whole experimental time horizon and to consider all measurement times jointly. This approach enables us to derive modification effects from estimated model parameters. We demonstrate that the exponential and logistic growth model recognise simulated scenarios more reliably than a statistical test.
Moreover, we apply the population growth models to evaluate the efficacy of candidate gene knockouts in patient-derived xenograft models of acute leukaemia.

\section*{Author summary}
In many areas of biomedical research, scientists study how diseases develop and respond to interventions over time. In leukaemia research, for example, genes are modified in cancer cells to test whether tumour growth can be decelerated. These experiments include repeated measurements over time to monitor disease progression. In practice, however, such data are frequently evaluated using standard statistical tests that compare selected measurements without explicitly accounting for the underlying dynamics. We argue that model-based evaluation of the complete time course provides a more informative and biologically grounded interpretation by capturing how the disease evolves over time.

In our study, we use mathematical descriptions of population growth to analyse all measurements together and to interpret them in light of underlying biological mechanisms. This allows us to move beyond asking whether a difference exists at a specific time point and instead to explore how a genetic modification changes the dynamics of disease progression. Using both simulated and experimental data from mouse models of acute leukaemia, we show that this strategy can more reliably detect meaningful effects.


\section*{Introduction}
Ordinary differential equations (ODEs) are suitable to mathematically represent mechanistic knowledge about processes of cells and biological systems in general~\cite{Banga2025}. This model-based approach allows researchers to validate their understanding of the system of interest and to study its characteristics for time-resolved data by inferring model parameters. Model parameters also allow them to quantify effects of interventions in the system.

An important area where ODE modelling shows potential is cancer research, for instance, in drug development. A major class of targeted cancer drugs is characterised by reducing the net growth of cancer cell populations by either inhibiting proliferation (thus cell division) or stimulating cell death for malignant cells. Developing drugs in this class is particularly challenging, especially in the early stages of research. Many therapeutic targets are explored, but they must pass several phases before they can reach clinical application. One crucial stage is preclinical trials. Here, relevant data about cell population growth-inhibiting effects is collected, including trials with laboratory animals.

\begin{figure}[!h]
	\centering
	\includegraphics[width=\textwidth]{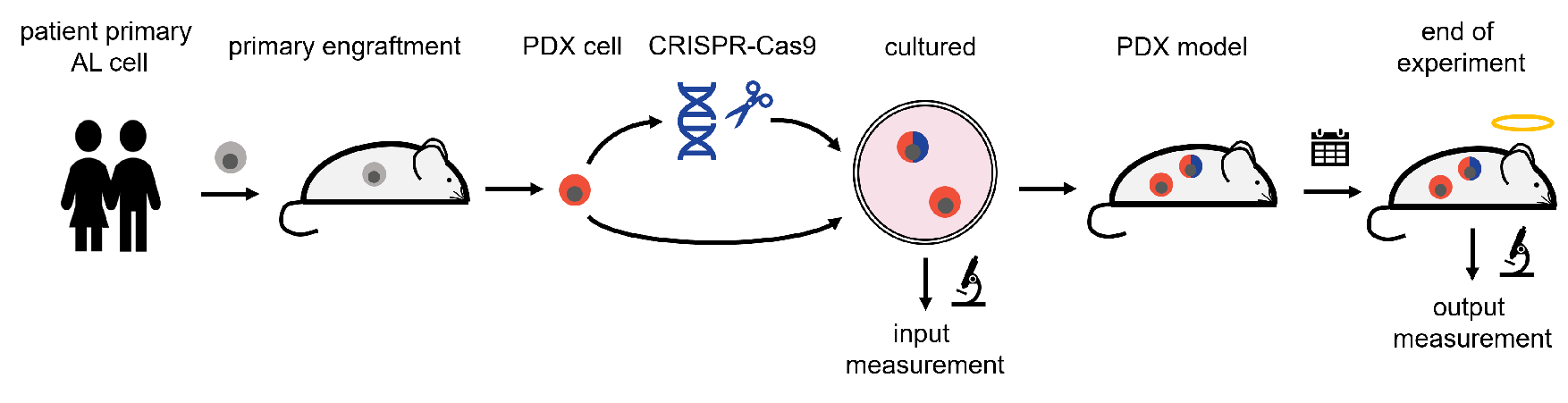}
	\caption{\textbf{Workflow of a gene modification experiment.}
    Schematic workflow of a CRISPR-Cas9 knockout experiment exploring new therapeutic targets for acute leukaemia (AL). First, patient cells are engrafted into mice to generate PDX cells. These cells are genetically modified and again engrafted into mice together with an untreated cell population providing PDX models. The output measurements, i.\,e.\ the concentrations of the modified population in the bone marrow of mice at an advanced stage of leukaemia, are compared with the input measurements, i.\,e.\ the concentration that is injected, to assess a possible growth-inhibiting effect on the modified cell population.}
	\label{biology:fig_experiment}
\end{figure}

In this work, we focus on preclinical experiments testing potential new therapeutic targets for acute leukaemia. Even though the mortality rate of acute leukaemia patients has decreased substantially over the past decades, leukaemia is still one of the main causes of death in children~\cite{Siegel2023}. To identify therapeutic targets, primary patient cells are engrafted into immune compromised NOD scid gamma (NSG) mice (NOD mice were originally used to study type 1 diabetes). After successful engraftment, patient-derived xenograft (PDX) cells are isolated from mice and can be genetically modified using, for example, the CRISPR-Cas9 (clustered regularly interspaced short palindromic repeats and CRISPR-associated protein 9) method~\cite{Wang2014,Shalem2014}. In particular, certain genes that are suspected to be relevant for proliferation or survival are knocked out~\cite{Bahrami2023,Ghalandary2023}. These cells are engrafted into NSG mice together with an untreated control population. Such PDX models constitute mouse models in which tumour cells can be studied in the environment of a living organism. Compared to cell cultures, these models better represent the growth environment of tumours in humans~\cite{Hidalgo2014,Gao2014}. At an advanced stage with a high blast load, a mouse needs to be sacrificed to measure the concentration of modified leukaemia cells in the bone marrow. This time point is sample-specific and may vary for different leukaemia samples used in the experiments. The output measurements are compared to the input concentrations, i.\,e.\ the proportions of the engrafted modified cells, to examine possible growth-inhibiting effects of the gene modification. An effective gene modification can have a proliferation-inhibiting or a cell death-stimulating effect. The essential steps of this kind of experiments are illustrated in Fig~\ref{biology:fig_experiment} in a simplified manner.

The standard approach to compare the input and output relations is to apply a statistical hypothesis test, e.\,g.\ a paired $t$-test, to assess whether the measurements at these two time points differ significantly from each other. This approach neglects mechanistic knowledge about cell reactions, such as proliferation, death and interactions, that are potentially affected by gene modifications and on the basis of which effects can be assessed. Moreover, this approach only compares output measurements from one time point with the corresponding inputs from day~$0$. The experiments on which this work is based, however, have up to two measurement times for output values. Therefore, up to two $t$-tests are required to evaluate them separately, although the measurements share the same experimental conditions (with different output times). This separation of evaluations induces small sample sizes for single paired $t$-tests, and so a sufficient number of measurements is required to detect a possible effect of a gene modification. As one mouse needs to be sacrificed for every single output measurement, researchers face a trade-off between the information content of experiments and ethical and financial considerations. 
Thus, the paired $t$-test approach ends up compounding challenges for efficient evaluation of preclinical leukaemia trials.

In this work, we provide a model-based approach for the evaluation of gene modification experiments on leukaemia cells that overcomes several limitations of the standard testing approach. We mathematically represent the series of cell events using ODE models to describe the dynamics of cell populations. In particular, we apply two common population growth models, an exponential and a logistic growth model~\cite{Wodarz2014}. A population growth model is incorporated in the recently developed algorithm Chronos~\cite{Dempster2021} to evaluate CRISPR-Cas9 knockout screens. Chronos describes single-guide RNA (sgRNA) depletion using an exponential growth model. It has outperformed other common algorithms in the evaluation of knockout screens, such as MAGeCK~\cite{Li2014,Li2015} and CERES~\cite{Meyers2017}, especially when time-series data is available~\cite{Dempster2021}. However, exponential growth of a population is often unrealistic in the long run, as it implies a constant doubling time and that population sizes either tend towards infinity or zero~\cite{Wodarz2014}. Therefore, we compare its performance with a logistic growth model involving an upper limit for population sizes. Logistic growth models have already been applied successfully to describe the growth of leukaemia cell populations~\cite{Ebinger2016,Hoffmann2020,Chulian2022}. For both models, population dynamics are characterised by parameters. To estimate these parameters, we employ the set of all experimental measurements taken at several output times to capture population dynamics over the entire experimental time horizon. Furthermore, we investigate parameter identifiability and derive confidence intervals for relevant parameters by means of profile likelihoods~\cite{Raue2009,Raue2010}. Based on estimated parameters and confidence intervals, we investigate growth-inhibiting effects of gene modifications.

We show in a simulation study that the exponential and logistic growth models perform similarly well in the evaluation of various simulated scenarios. For small sample sizes and in the presence of a weak or absent effect, the exponential growth model yields more stable parameter estimation. A paired $t$-test for late output measurements can compete with the population growth models in detecting modification effects, at least for this specific allocation of measurement times of the considered type of experiments and for a sufficiently large sample size. For experimental data of gene modification experiments, the growth models and the paired $t$-test for late measurement times yield similar evaluation results. Nonetheless, the growth models provide a more informative approach to analyse underlying dynamics, as they explicitly account for the temporal structure of the data and allow mechanistic interpretation of biologically meaningful parameters.

\section*{Methods}
\subsection*{Dynamic growth models}

Biological systems evolving over time are often mathematically described by ODEs. We consider two-dimensional ODE systems for the gene modification experiments with state variable vector $\bm{x}(t)=(x_1(t),x_2(t))^{\mathrm T}$ representing the leukaemia cell numbers of both the modified ($x_1(t)$) and the untreated ($x_2(t)$) population at time $t\ge0$. At any time, cells of both populations can perform one out of two reactions: they can either grow and divide (proliferation) or die (apoptosis). These two possible reactions determine the dynamics of population numbers. To describe the dynamics formally, we examine two growth models that are both embedded into the broad class of ODEs defined by 
\begin{align}\label{models:ode_general}
	\mathrm{d} \bm{x}(t)=\bm{f}(\bm{x}(t),\bm{\theta})\mathrm{d} t,\quad \bm{x}(0)\in\mathbb{R}^2.
\end{align}
The concrete choice of the $\mathbb{R}^2$-valued function~$\bm{f}$ depends on the considered growth model. The function~$\bm{f}$ is parametrised by a parameter vector $\bm{\theta}\in\mathbb{R}^p$, the value of which is either specified by the modeller, or it is estimated from data. The vector~$\bm{\theta}$ may also contain the initial conditions $\bm{x}(0)=(x_1(0),x_2(0))^{\mathrm T}$ of the state variable~$\bm{x}$ if these are unknown.

The state variable components $x_k(t)$, $k=1,2$, are $\mathbb{R}$-valued quantities, whereas the number of cells for the modification experiments are integers. In the application considered here, cell numbers range from $10^4$ to $10^9$. These scales can be regarded as sufficiently large to treat the trajectory of cell numbers as approximately continuous on a macroscopic level such that an ODE representation is appropriate to impose. 

In the following, we introduce an exponential and a logistic growth model to characterise the growth of the two cell populations within the gene modification experiments. These two models are widely used to mathematically describe the growth of cancer~\cite{Wodarz2014}, as well as the growth of populations in general~\cite{Kot2001}. Moreover, we define the observable model that represents the form of the experimental measurements. We simulated both the trajectories of the ODE systems and the observables with the simulation framework of the~\textsf{dMod} package (Version 1.0.2) in \textsf{R}~\cite{Kaschek2019}.

\subsubsection*{Exponential growth model}

\begin{figure}[!h]
	\centering
	\includegraphics[width=0.8\textwidth]{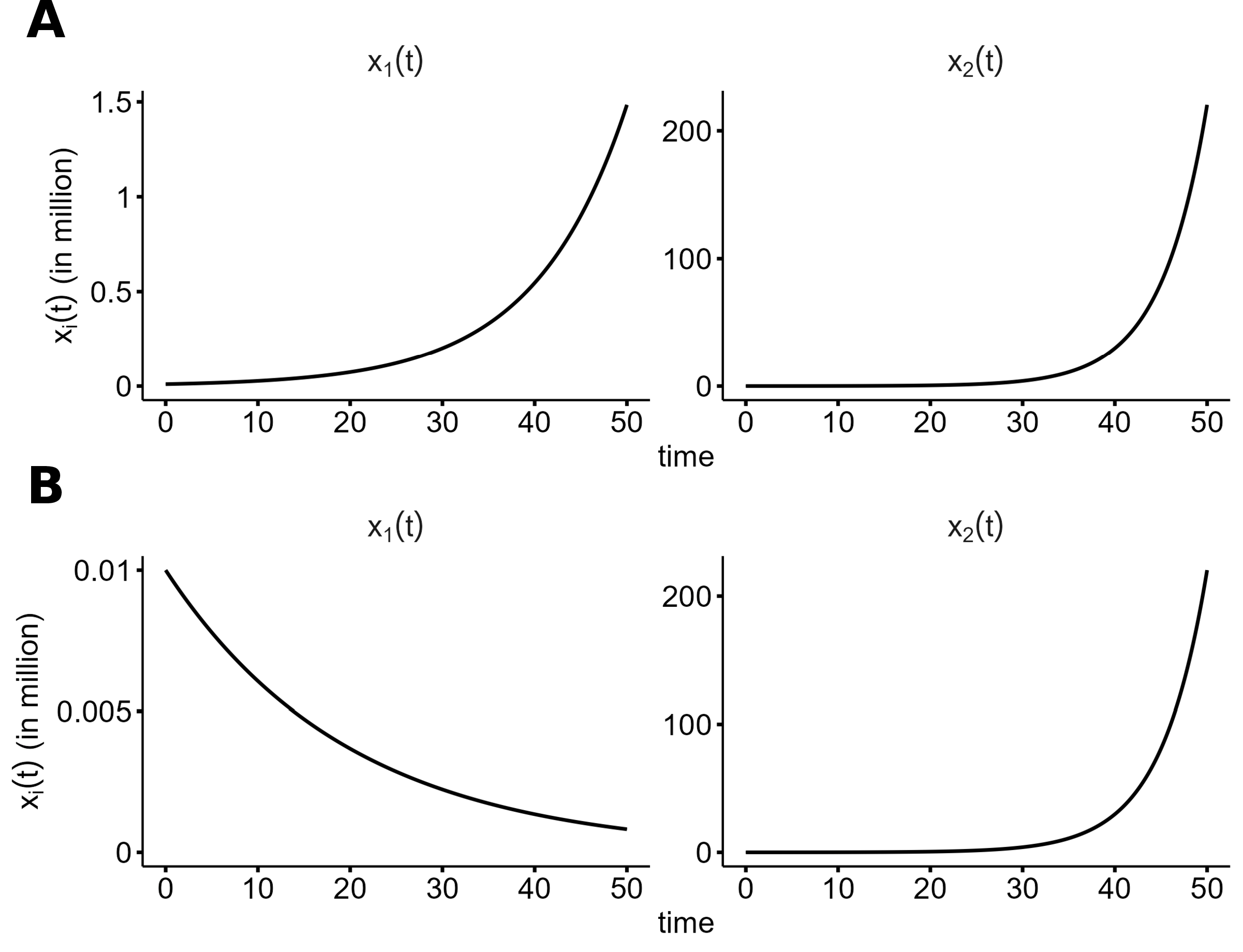}
	\caption{\textbf{Two scenarios for the exponential growth model.}
    Trajectories of the exponential growth model with exemplary values for $\beta_1$ and $\beta_2$ with $x_1(0)=x_2(0)=10^4$. (A) Both populations grow, as $\beta_1=0.1>0$ and $\beta_2=0.2>0$, but Population~1 grows less strongly than Population~2. (B) Population~1 shrinks and Population~2 grows, as $\beta_1=-0.05<0$ and $\beta_2=0.2>0$.}
	\label{models:fig_exp}
\end{figure}

We assume that the per capita net growth rate of cells from Population~$k$ is given by the parameter $\beta_k\in\mathbb{R}$ for $k=1,2$. The parameter reflects the difference of the per capita proliferation rate and the death rate of the population. The resulting dynamics of the cell population numbers are represented by the ODE system
\begin{align}\label{models:ode_exp}
	\begin{split}
		\mathrm{d} x_1(t)&=\beta_1 x_1(t)\mathrm{d} t,\quad x_1(0)\in\mathbb{R},\\
		\mathrm{d} x_2(t)&=\beta_2 x_2(t)\mathrm{d} t,\quad x_2(0)\in\mathbb{R}.
	\end{split}
\end{align}
The system~\eqref{models:ode_exp} admits the analytic solution
\begin{align}\label{models:ode_exp_solution}
		x_1(t)=x_1(0)\mathrm{e}^{\beta_1t},\quad
		x_2(t)=x_2(0)\mathrm{e}^{\beta_2t},\quad t\ge0.
\end{align}
The solution in Eq~\eqref{models:ode_exp_solution} shows that, if~$\beta_k>0$, Population~$k$ grows exponentially and, if~$\beta_k<0$, it declines exponentially. In the application considered in this article, scientists seek after gene modifications that make Population~1 grow less strongly or even shrink in comparison to Population~2, i.\,e.\ $\beta_1<\beta_2$. These two desirable cases are depicted in Fig~\ref{models:fig_exp} for exemplary parameter values.

We assume that a given gene modification exerts the same effect on the same leukaemia sample across different mice; that is, different mice share identical growth rates~$\beta_k$. To account for inter-mouse heterogeneity and variability in cell homing~\cite{Ebinger2016,Bahrami2023}, we allow for mouse-specific initial conditions~$x_k(0)$. For the sake of notational simplicity, we omit this dependence in the notation of initial conditions.

The exponential nature in Eq~\eqref{models:ode_exp_solution} implies that $x_k(t)\to\infty$ if $\beta_k>0$, or $x_k(t)\to0$ if $\beta_k<0$, as $t\to\infty$. Both this asymptotic behaviour and a constant doubling time are rarely observed in real phenomena in the long run. At some point, for example, food resources or the space in the system are exhausted such that growth decelerates~\cite{Wodarz2014}. In our application, however, we are more interested in the evaluation of population dynamics on a relatively small time horizon than in predicting beyond this horizon. It has been observed that exponential growth of a population in a limited time period can be reasonably assumed, in particular for non-solid tumours as leukaemia~\cite{Simon2006,Wodarz2014}. Thus, it still makes sense to employ the exponential growth model. In addition, however, we will investigate the logistic growth model which takes into account an upper bound for population sizes.

\subsubsection*{Logistic growth model}

\begin{figure}[!h]
	\centering
	\includegraphics[width=0.8\textwidth]{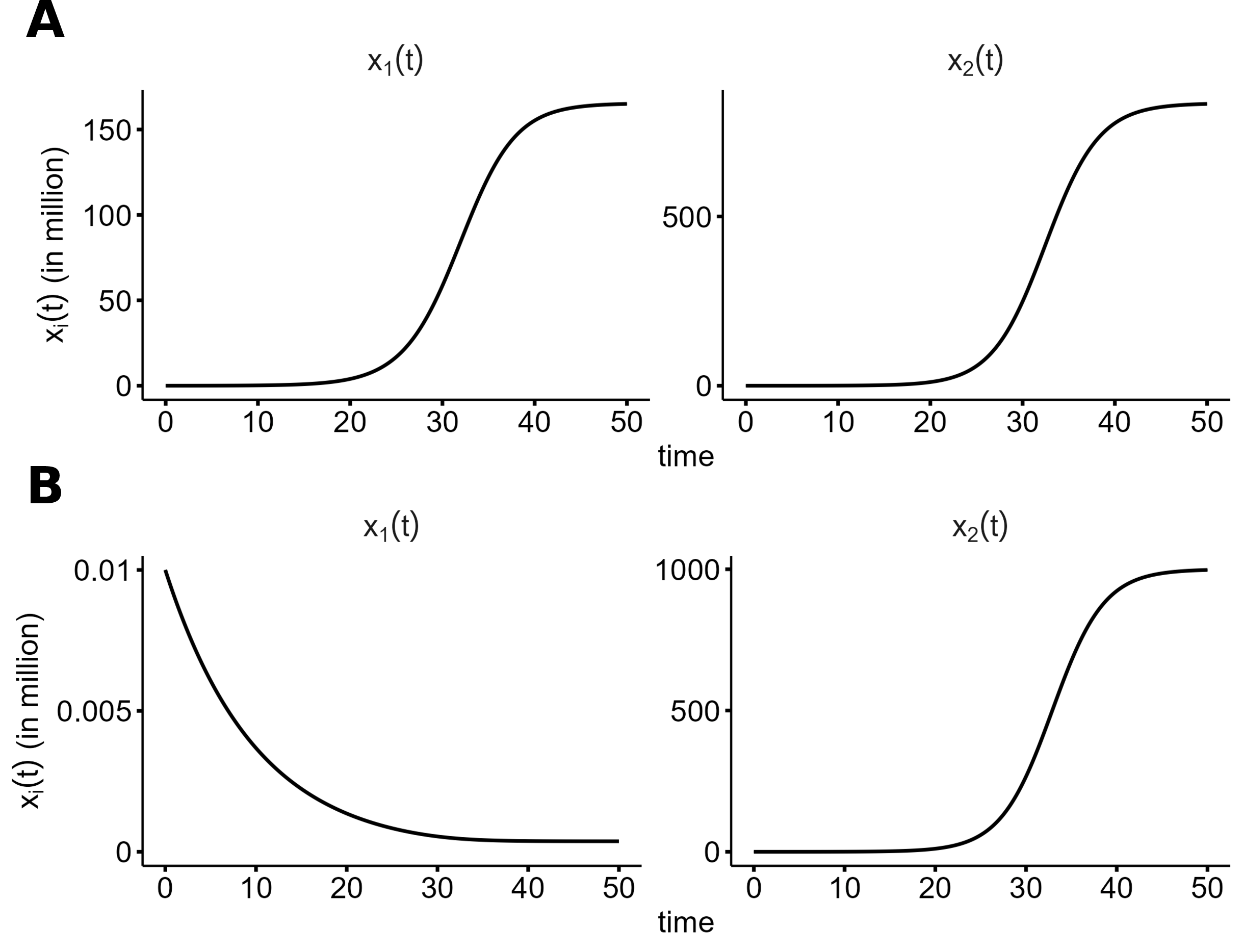}
	\caption{\textbf{Two scenarios for the logistic growth model.}
    Trajectories of the logistic growth model with exemplary values for $\lambda_1$ and $\lambda_2$ with $K=10^9$ and $x_1(0)=x_2(0)=10^4$. (A) Both populations grow, as $\lambda_1=0.3>0$ and $\lambda_2=0.35>0$, but Population~1 grows less strongly than Population~2. (B) Population~1 shrinks and Population~2 grows, as $\lambda_1=-0.1<0$ and $\lambda_2=0.35>0$.}
	\label{models:fig_logistic}
\end{figure}

A logistic growth model accounts for the fact that real-world populations rarely grow permanently exponentially. It assumes that the sum of all system entities cannot exceed a carrying capacity $K\in\mathbb{N}$ at any time~$t$ if the sum of the initial conditions is less than $K$, i.\,e.\ $x_1(0)+x_2(0)<K$ in our case. In the considered context, the capacity~$K$ reflects the finite resources for leukaemia cells to proliferate in the bone marrow of mice and we presume that the bone marrow can only absorb a maximum of $K$ leukaemia cells. The corresponding logistic growth model reads
\begin{align}\label{models:ode_logistic}
	\begin{split}
		\mathrm{d} x_1(t)&=\lambda_1\left(1-\frac{x_1(t)+x_2(t)}{K}\right)x_1(t)\mathrm{d} t,\quad x_1(0)\in\mathbb{R},\\
		\mathrm{d} x_2(t)&=\lambda_2\left(1-\frac{x_1(t)+x_2(t)}{K}\right)x_2(t)\mathrm{d} t,\quad x_2(0)\in\mathbb{R},
	\end{split}
\end{align}
for net growth rates $\lambda_1,\lambda_2\in\mathbb{R}$ of Population~1 and Population~2, respectively. As the initial conditions usually satisfy $x_1(0)+x_2(0)<K$, Population~$k$ grows if $\lambda_k>0$, and, if $\lambda_k<0$, the population size declines. If $x_1(t)+x_2(t)\ll K$ for small $t$, the term in the brackets in~\eqref{models:ode_logistic} is close to one and the number of cells grows (or declines) almost exponentially. The closer the sum $x_1(t)+x_2(t)$ approaches the threshold $K$ for larger $t$, the stronger the growth (or decay) decelerates until the system reaches a steady state. The value of the steady state depends on the values of the parameters $\lambda_1$, $\lambda_2$ and~$K$ and the initial conditions~$x_1(0)$ and~$x_2(0)$. So, the shared capacity links the behaviour of both cell populations, and the growth of both populations depends on the other's growth activity. An analytically explicit solution like~\eqref{models:ode_exp_solution} is unavailable for~\eqref{models:ode_logistic}. For parameter values of a less strongly growing as well as a shrinking Population~1, simulated model trajectories are shown in Fig~\ref{models:fig_logistic}. Again, we consider mouse-specific initial conditions~$x_k(0)$ for this growth model in subsequent sections and omit this dependence notation-wise.

\subsubsection*{Observable model}

\begin{figure}[!h]
	\centering
	\includegraphics[width=0.8\textwidth]{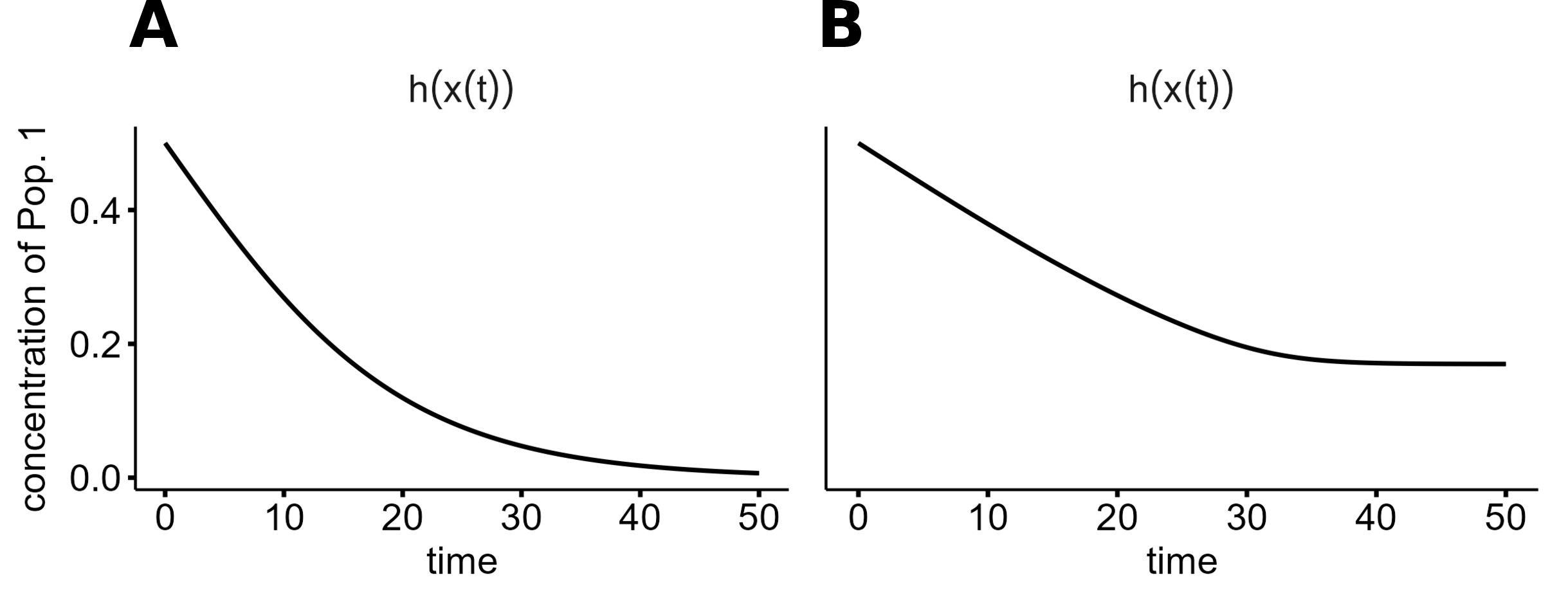}
	\caption{\textbf{Two scenarios for the observable.}
    Trajectories of the observable~$h$ for (A) the exponential and (B) the logistic growth model. The parameter values correspond to those in Fig~\ref{models:fig_exp}A and Fig~\ref{models:fig_logistic}A.}
	\label{models:fig_observable}
\end{figure}

For gene modification experiments, it is technically difficult to measure absolute population sizes, i.\,e.\ $x_1(t)$ and $x_2(t)$, explicitly. Therefore, it is common to measure the concentration of one population in relation to the total number of cells of both populations. The observable for both growth models thus results as 
\begin{align}\label{models:observable}
	h(\bm x(t))&=\frac{x_1(t)}{x_1(t)+x_2(t)},\quad t\ge0.
\end{align}
For the exponential growth model~\eqref{models:ode_exp}, we can rewrite Eq~\eqref{models:observable} and plug in \eqref{models:ode_exp_solution} such that we obtain
\begin{align}\label{app:obs_exp_rewritten}
	h(t)
	=\frac{1}{1+\frac{x_2(t)}{x_1(t)}}
	=\frac{1}{1+\frac{x_2(0)}{x_1(0)}\mathrm{e}^{(\beta_2-\beta_1)t}},\quad t\ge0.
\end{align}
Trajectories of the observable are depicted in Fig~\ref{models:fig_observable} for the dynamics of the exponential and the logistic growth model from Fig~\ref{models:fig_exp}A and Fig~\ref{models:fig_logistic}A, respectively. In Fig~\ref{models:fig_observable}A, the relative amount of Population~1 tends to zero due to the exponential nature of the model, whereas the observable reaches a steady state around~$0.15$ in the long run in Fig~\ref{models:fig_observable}B.

\subsection*{Statistical inference}

A mathematical model attempts to depict real mechanisms so that insights can be transferred from the model to reality. The values of model parameters can sometimes be determined by the modeller on the basis of knowledge from the specific discipline. In the case of the gene modification experiments considered here, neither the parameter values nor the initial conditions are exactly known. The goal of statistical inference is to estimate these values based on measurements and to study resulting properties. However, models are only an idealized representation of reality. Measurements naturally deviate from the modelled observable~$h$ as they are, for example, subject to biological stochasticity and experimental noise. We regard measurements~$y(t)$ for both growth models as realisations of the random variable
\begin{align}\label{statistics:observable_error}
	Y(t)=h(\bm x(t))+\varepsilon(t)
	=\frac{x_1(t)}{x_1(t)+x_2(t)}+\varepsilon(t),\quad t\ge0,
\end{align}
where $\varepsilon(t)\sim\mathcal{N}\left(0,\sigma^2\right)$ represents normally distributed measurement error, independent and identically distributed for all~$t$, with standard deviation $\sigma>0$. 

For both the exponential and the logistic growth model, we combine all possible unknown model and error parameters into the parameter vectors $\bm{\theta}^{(e)}\in\Theta_j^{(e)}$ and $\bm{\theta}^{(l)}\in\Theta_j^{(l)}$, respectively. The dimension of these vectors may vary across experiments depending on the number of mice used and, consequently, on the number of mouse-specific initial conditions.

The statistical methods described in the following can be applied for both models. For the remainder of this section, we, therefore, use $\bm{\theta}\in\Theta$ as a generic parameter vector, representing both $\bm{\theta}^{(e)}$ and~$\bm{\theta}^{(l)}$. We let $\bm{y}=(y_1,\dots,y_n)$ denote a data set with~$n$ measurements taken at time points $0\le t_1\le\dots\le t_n$. The measurements of~$\bm{y}$ are assumed to be independent realisations of the random variable~$Y$ from Eq~\eqref{statistics:observable_error} at the respective time points. The log-likelihood function for~$\bm{y}$ thus reads
\begin{align}\label{statistics:likelihood}
	\ell(\bm y|\bm{\theta})=\sum_{i=1}^n\log \phi\left(y_i\left|h(\bm{x}(t_i;\bm{\theta})),\sigma^2 \right.\right),
\end{align} 
where $\phi\left(\cdot\left|h(\bm{x}(t_i;\bm{\theta})),\sigma^2\right.\right)$ is the density function of the normal distribution with mean $h(\bm{x}(t_i;\bm{\theta}))$ and standard deviation~$\sigma$. The maximum likelihood estimate of~$\bm{\theta}$ is given by 
\begin{align*}
	\hat{\bm{\theta}}
	=\underset{\bm{\theta}\in\Theta}{\arg\max}~\ell(\bm y|\bm{\theta}).
\end{align*} 

To further infer the components of~$\bm{\theta}$ based on a data set~$\bm{y}$, we introduce profile likelihoods to calculate confidence intervals and to investigate parameter identifiability. These methods are applied in subsequent sections to infer the model parameters for simulated scenarios as well as for knockout experiments in PDX models of acute leukaemia.

\subsubsection*{Profile likelihood}

Since observations are subject to noise, parameters are estimated with uncertainty. To assess the precision of an estimate, we calculate confidence intervals for the components of~$\bm{\theta}$ based on profile likelihoods. The profile likelihood of a parameter~$\theta_j$ is defined by
\begin{align*}
	\text{PL}_j(q)=\max_{\{\bm{\theta}|\theta_j=q\}}\ell(\bm y|\bm{\theta}),\quad q\in\Theta_j,
\end{align*}
where~$\Theta_j$ is the domain of~$\theta_j$~\cite{Raue2009}. To compute~$\text{PL}_j(q)$, the log-likelihood is maximised with respect to all components~$\theta_l$, $l\neq j$, keeping~$\theta_j=q$ fixed. By evaluating~$\text{PL}_j(q)$ for a range of values~$q$ around the component~$\hat{\theta}_j$ of the maximum likelihood estimate~$\hat{\bm{\theta}}$, the shape of the likelihood along the axis of~$\theta_j$ is explored. An approximate likelihood-based confidence interval for $\theta_j$ at confidence level~$1-\alpha$ is then given by 
\begin{align}\label{statistics:prof_lik_ci}
	\text{CI}_{1-\alpha}^{\chi_1^2}(\theta_j)=\left\{q\left|2\cdot\left(\ell(\bm y|\hat{\bm\theta})-\text{PL}_j(q)\right)\le\Delta_{\alpha}^{\chi_1^2}\right.\right\},
\end{align}
where $\Delta_{\alpha}^{\chi_1^2}$ denotes the $(1-\alpha)$-quantile of the $\chi^2$-distribution with one degree of freedom (in short: $\chi_1^2$-distribution), for $\alpha\in(0,1)$~\cite{Raue2010}. A confidence interval~$\text{CI}_{1-\alpha}^{\chi_1^2}(\theta_j)$ consists of the union of disjoint intervals, if the difference of log-likelihoods in Eq~\eqref{statistics:prof_lik_ci} exceeds the quantile more than twice. 

In~\cite{Toensing2023}, the authors show that the type of confidence interval from Eq~\eqref{statistics:prof_lik_ci}, assuming an asymptotic setting with the $\chi_1^2$-distribution, often inaccurately reflects parameter uncertainty for dynamic models in finite-sample cases. We investigate the appropriateness of this asymptotic assumption for both growth models within a subsection of the analysis of gene knockout trials in PDX models.

In the following, we show how the confidence intervals defined in \eqref{statistics:prof_lik_ci} are used to check parameter identifiability. Later, we introduce evaluation methods of the gene modification experiments for the growth models relying on confidence intervals of growth parameters.

\subsubsection*{Parameter identifiability}

Before we compute and interpret parameter estimates, we check whether the model parameters can be unambiguously identified. The literature distinguishes between two types of parameter identifiability: a parameter~$\theta_j$ is called structurally identifiable if a unique maximum of~$\ell(\bm y|\bm{\theta})$ exists with respect to~$\theta_j$. Otherwise, the parameter~$\theta_j$ is structurally non-identifiable. In this case, $\text{PL}_j(q)$ is, in general, constant, which means that the model parametrisation is ambiguous in the sense that different parameter values yield the same model trajectories of $\bm{x}(t)$ or $h(\bm x(t))$~\cite{Raue2010}. Structural non-identifiability can usually be resolved by a model reparametrisation or fixing parameters to certain values~\cite{Wieland2021}. A more detailed discussion on structural identifiability is given in~\cite{Anstett-Collin2020} which further distinguishes between local and global identifiability.

Even if a parameter~$\theta_j$ is structurally identifiable, insufficient data may still hamper to determine its value with finite confidence bounds. In this case, the difference of log-likelihoods only exceeds the quantile in Eq~\eqref{statistics:prof_lik_ci} in one direction at most. Therefore, parameter~$\theta_j$ is said to be practically identifiable if the confidence interval~$\text{CI}_{1-\alpha}^{\chi_1^2}(\theta_j)$ is finite, for a confidence level~$1-\alpha$~\cite{Raue2010}. To make a practically non-identifiable but structurally identifiable parameter identifiable, the data quality or quantity must be increased to narrow its confidence interval~\cite{Wieland2021}.

\subsection*{Ethics statement}
For the CRISPR–Cas9-based knockout experiments presented in the subsequent section, written consent forms were obtained from all patients and from parents/carers in cases where patients were minors. The study was performed following the ethical standards of the responsible committee on human experimentation (written approval by Ethikkommission des Klinikums der Ludwig-Maximilians-Universit\"at M\"unchen, ethikkommission@med.uni-muenchen.de, April 15/2008, number 068-08, 222-10) and with the Helsinki Declaration of 1975, as revised in 2013.
All animal trials were performed by the current ethical standards of the official committee on animal experimentation written approval by Regierung von Oberbayern, ROB-55.2Vet-2532.Vet\_02-15-193, ROB-55.2Vet-2532.Vet\_03-16-56, ROB-55.2-2532.Vet\_02-20-221 and ROB-55.2Vet-2532.Vet\_02-16-7, ROB-55.2-2532.Vet\_02-20-159, ROB-55.2-2532.Vet\_0321-9, ROB-55.2-2532.Vet\_02-23-78).

\section*{Results}
\subsection*{Evaluation of effect detection performance for simulated data}

In a simulation study, we investigated whether incorporating domain-specific knowledge about the data generating process makes the analysis of time-resolved data more informative than a statistical test. We simulated data from a stochastic process relating to different growth-inhibiting scenarios of leukaemia cell populations. Sample sizes and effect sizes were motivated by the data from knockout experiments in PDX models used in our study. We represented the process by an exponential and logistic growth model and compared their detection performance with paired $t$-tests in various scenarios.  

Before we introduce the evaluation methods and simulation scenarios, we check structural identifiability of the two growth models. For this section as well as for the analysis of gene knockout trials in PDX models, we applied the \textsf{dMod} package in \textsf{R} to estimate parameters and to calculate profile likelihood-based confidence intervals. For this purpose, we added an $L_2$-constraint on the parameters to the objective function such that the trust region optimiser avoided selecting extreme parameter values of a possibly non-identifiable model. We removed the constraint for assessing identifiability and for computing confidence intervals as these concepts require data-based results~\cite{Kaschek2019,Riesle2023}.

\subsubsection*{Structural identifiability of the growth models}

We investigated structural identifiability of the parameters of both the exponential and the logistic growth model using profile likelihoods. For this analysis, we used simulated data of comparably large sample size. We computed one trajectory for each ODE model for one concrete parameter setting using the simulation method from the \textsf{dMod} package. We extracted the computed values for every integer time point between~$0$ and~$50$. We added realisations of a normally distributed random variable to each simulated value to obtain realisations of the random variable~$Y$ from Eq~\eqref{statistics:observable_error}. By this means, we simulated~$20$ observations for each time point yielding one data set for each growth model, each containing~$1020$ observations. In this theoretical setting, we worked with a single pair of initial conditions.

The exponential growth model exhibited the parameter vector $\bm{\theta}^{(e)}=(\beta_1,\beta_2,x_1(0),x_2(0),\sigma)^{\mathrm T}$ and we set $\bm{\theta}^{(e)}=(0.3,0.5,50,50,0.05)^{\mathrm T}$ for the data generation. Since $x_1(0)$, $x_2(0)$ and $\sigma$ were assumed to be non-negative, we considered log-transforms of these components during numerical optimisation; this way, we could apply the trust region optimiser \textsf{trust} from \textsf{dMod} for performing unconstrained optimisation. The profile likelihoods of~$\bm{\theta}^{(e)}$ revealed that all components except~$\sigma$ were structurally non-identifiable as the negative profile likelihoods exhibited no unique minimum (Fig~\ref{app:fig_exp_model_non-ident_model}). This finding was hardly surprising taking into account how the parameters are related within the closed-form expression of the observable function in Eq~\eqref{app:obs_exp_rewritten}: the initial conditions are contained as a ratio and the growth parameters form the relative net growth of both populations. Therefore, we reparametrised the model with $\bm{\theta}^{(e)}=(\theta_1^{(e)},\theta_2^{(e)},\theta_3^{(e)})^{\mathrm T}$ by setting
\begin{align*}
	\theta_1^{(e)}=\beta_2-\beta_1,\quad \theta_2^{(e)}=\frac{x_2(0)}{x_1(0)},\quad \theta_3^{(e)}=\sigma,
\end{align*}
while retaining the name of the overall parameter. Again, we log-transformed~$\theta_2^{(e)}$ and~$\theta_3^{(e)}$ for numerical unconstrained optimization. All components of~$\bm{\theta}^{(e)}$ turned out to be structurally identifiable (Fig~\ref{app:fig_exp_model_ident_model}). In the remainder of this work, we treat the parameter~$\theta_2^{(e)}$ as mouse-specific.

For the logistic growth model and its parameter vector $\bm{\theta}^{(l)}=(\lambda_1,\lambda_2,K,x_1(0),x_2(0),\sigma)^{\mathrm T}$, we set $\bm{\theta}^{(l)}=(0.1,0.3,1000,50,50,0.05)^{\mathrm T}$ for the data simulation. We log-transformed the parameters~$K$, $x_1(0)$, $x_2(0)$, and $\sigma$. The profile likelihoods of the components of~$\bm{\theta}^{(l)}$ showed that $K$, $x_1(0)$ and $x_2(0)$ were structurally non-identifiable (Fig~\ref{app:fig_log_model_non-ident_model}). When we set $x_1(0)=50$ to the true value of the data generation, we obtained a structurally identifiable model (Fig~\ref{app:fig_log_model_ident_model}). However, if the growth parameters~$\lambda_1$ and~$\lambda_2$ were equal, e.\,g.\ $\lambda_1=\lambda_2=0.1$, both parameters would be structurally non-identifiable, even if~$x_1(0)$ was fixed.

\subsubsection*{Evaluation methods}

The parameters~$\theta_1^{(e)}=\beta_2-\beta_1$, $\theta_1^{(l)}=\lambda_1$ and~$\theta_2^{(l)}=\lambda_2$ described the growth behaviour of the two leukaemia cell populations within the exponential and the logistic growth model, respectively. Based on confidence intervals related to these parameters, we evaluated simulated experimental scenarios as well as real-world experiments in PDX models of acute leukaemia. 

The parameter $\theta_1^{(e)}$ of the exponential growth model reflected the net growth of the modified Population~1 in relation to Population~2. If the confidence interval $\text{CI}_{1-\alpha}^{\chi_1^2}(\theta_1^{(e)})$, for $\alpha\in(0,1)$, excluded zero, we deduced that the growth behaviour of Population~1 differed significantly from that of Population~2. This was equivalent to a two-tailed significance test with null hypothesis $H_0:\theta_1^{(e)}=0$ and alternative hypothesis $H_1:\theta_1^{(e)}\neq0$ at significance level~$\alpha$. We found in a subsequent verification that the approximate confidence interval $\text{CI}_{1-\alpha}^{\chi_1^2}(\theta_1^{(e)})$, as defined in Eq~\eqref{statistics:prof_lik_ci}, tended to underestimate parameter uncertainty for knockout experiments in PDX models using quantiles of the $\chi_1^2$-distribution. For a confidence level of 95\,\%, we therefore calculated confidence intervals $\text{CI}_{0.95}^{\text{Cant}}(\theta_1^{(e)})$ using the adapted threshold $\Delta_{0.05}^{\text{Cant}}=7.16$ in addition to using the quantile $\Delta_{0.05}^{\chi_1^2}=3.84$ of the $\chi_1^2$-distribution, as suggested in~\cite{Toensing2023}.

For the logistic growth model, the parameters $\theta_1^{(l)}=\lambda_1$ and $\theta_2^{(l)}=\lambda_2$ represented the net growth of Population~1 and Population~2, respectively. We assessed differences in growth behaviour by considering their difference $\delta=\lambda_2-\lambda_1$, which quantified deviations from identical growth dynamics. To directly estimate~$\delta$ within the model formulation, we reparametrised $\lambda_2=\delta+\lambda_1$ in Eq~\eqref{models:ode_logistic}. 
This reparametrisation was structurally identifiable (cf.\ Fig~\ref{app:fig_log_model_ident_model_delta}) and enabled computation of a profile likelihood-based confidence interval $\text{CI}_{1-\alpha}^{\chi_1^2}(\delta)$. We evaluated whether this interval excluded zero, corresponding to the significance test with hypotheses $H_0:\delta=0$ vs. $H_1:\delta\neq0$ at significance level~$\alpha$. As detailed later, we additionally constructed confidence intervals $\text{CI}_{0.95}^{\text{Cant}}(\delta)$ using the adapted threshold $\Delta_{0.05}^{\text{Cant}}=7.16$ for the logistic growth model for $\alpha=0.05$.

The standard evaluation approach applied paired $t$-tests to compare measurements from only two different time points per experiment. The test assessed whether the mean difference between the paired input and output measurements differed significantly from zero. The test statistic was given by
\begin{align*}
	\tau=\frac{\frac{1}{m}\sum_{j=1}^m(y_j(0)-y_j(t_{\text{out}}))}{\frac{\sigma_D}{\sqrt{m}}},
\end{align*}
where $y_j(0)$ and $y_j(t_{\text{out}})$ denoted the input measurement from day~$0$ and output measurement from day~$t_{\text{out}}$ of mouse~$j$, respectively, and $\sigma_D$ denoted the empirical standard deviation of the differences between all $m$ mice considered. The corresponding hypotheses were $H_0:\tau=0$ vs.\  $H_1:\tau\neq0$ for significance level~$\alpha$. We conducted these $t$-tests without multiple testing correction, as the focus of our analysis was to individually compare the method of $t$-tests with model-based evaluations.

Although we were mainly interested in assessing whether a gene modification caused a cell population growth-inhibiting effect, i.\,e.\ $H_1:\theta_1^{(e)}>0$, $H_1:\delta>0$, or $H_1:\tau>0$, we related our evaluations to two-tailed significance tests; this was consistent with the standard procedure to analyse the experiments with two-tailed paired $t$-tests~\cite{Ghalandary2023}. We however retained to refer to the aim to recognise possible population growth-inhibiting effects.

\subsubsection*{Simulation study: data simulation with Gillespie's algorithm}

For the simulation study, we generated data sets mimicking scenarios with different effect and sample sizes with Gillespie's algorithm~\cite{Gillespie1976,Gillespie1977} using the \textsf{R} package \textsf{GillespieSSA} (Version 0.6.2,~\cite{Pineda-Krch2008}). Gillespie's algorithm simulates exact trajectories of a Markov jump process (MJP), i.\,e.\ a continuous-time Markov process with discrete states. Compared to a deterministic ODE representation that we have considered so far, an MJP represents the temporal sequence of biochemical reactions stochastically. It can be shown that the length of time intervals between two consecutive reactions is exponentially distributed for an MJP, which is utilised by Gillespie's algorithm. Due to the discrete nature and stochasticity of biological systems, MJPs and Gillespie's algorithm are commonly used to describe and simulate reaction networks, such as cell reactions in our case~\cite{Wilkinson2019}. However, the likelihood of an MJP is, in general, intractable which makes statistical inference challenging. Therefore, an ODE approximation is often used to represent reaction networks and to infer model parameters based on data.

Since Gillespie's original direct method is computationally intensive for large cell numbers, we applied the approximate optimised tau-leaping method~\cite{Cao2006} implemented in \textsf{GillespieSSA} to simulate trajectories of the state variable~$\bm{x}(t)$. For the data simulation, we assumed that a leukaemia cell of both populations could perform one out of two reactions at any time, it could either proliferate or die. These two reactions occurred with rates~$b_k$ and $d_k$ for Population~$k$, $k=1,2$. The propensity function for proliferation was given by $a_k^{(b)}(x)=b_k\cdot x_k$, while the propensity for cell death was $a_k^{(d)}=(d_k+(b_k-d_k)\cdot((x_1+x_2)/K)\cdot x_k$. The propensities for cell death ensured that cell death occurred more frequently with increasing population sizes, reflecting the limited capacity in the bone marrow~\cite{Pineda-Krch2008}. The more the sum of the population sizes approached the capacity~$K$, the more the propensities for death approached the proliferation rates, so that the net growth tended to zero for both populations. A deterministic approximation of the resulting MJP yielded the so-called reaction rate equation which equals the two-dimensional logistic growth model~\cite{Wilkinson2019}. 

We considered four scenarios with different parameter settings mimicking experiments with no gene modification effect, and with a small, medium and strong modification effect. We imposed that a gene modification only affected the cell death rate~$d_1$ of Population~1, while the proliferation activity of both populations and cell death of Population~2 remained unaffected. Therefore, only~$d_1$ varied across the scenarios, and the chosen values of the rate parameters are listed in Table~\ref{app:simulation_scenarios}. For all scenarios, we set $K=10^9$ and $x_1(0)=x_2(0)=5\cdot10^4$, and, to generate realisations of~$Y$ from Eq~\eqref{statistics:observable_error}, we set $\sigma=0.05$. Realisations of Eq~\eqref{statistics:observable_error} could take negative values due to the additive and normally distributed error and were manually set to zero accounting for the context.

\begin{table}[!h]
\centering
	\caption{\textbf{Simulated modification scenarios and corresponding parameter values.}}
	\label{app:simulation_scenarios}
	\begin{tabular}{lllll}
			\hline
			Scenario & $b_1$ & $d_1$ & $b_2$ & $d_2$ \\ 
			\hline
			no effect & $0.5$ & $0.1$ & $0.5$ & $0.1$ \\
			weak effect & $0.5$ & $0.12$ & $0.5$ & $0.1$ \\
			medium effect & $0.5$ & $0.15$ & $0.5$ & $0.1$ \\
			strong effect  & $0.5$ & $0.2$ & $0.5$ & $0.1$ \\			
			\hline
	\end{tabular}
    \begin{flushleft}
        Parameter values of the reaction rates to simulate four experimental scenarios with Gillespie's algorithm, where we further set $K=10^9$, $x_1(0)=x_2(0)=5\cdot10^4$, and $\sigma=0.05$ for all scenarios.
    \end{flushleft}
\end{table}

For each scenario, we simulated~$100$ data sets each mimicking a real experiment designed with two output measurement times, in this case day~$14$ and day~$40$. As each combination of one input (from day~$0$) and one output measurement (from day~$14$ or~$40$) corresponded to one mouse in knockout experiments in PDX models, we generated one Gillespie trajectory for each combination. We considered three sample size settings. In the first setting, we assumed that two fictitious mice were sacrificed and measured after $14$ days and two more after $40$ days, resulting in an overall sample size of eight measurements, i.\,e.\ four output measurements together with four input measurements from day~$0$. In the remaining two settings, we considered data sets with sample sizes of~$16$ and~$32$, where four and eight fictitious mice each were measured on day~$14$ and~$40$. In total, we thus simulated~$1200$ data sets. 

Each simulated Gillespie trajectory was a realisation of a stochastic process and could be regarded as the individual response of a mouse to a specific gene modification. This motivated us to consider mouse-specific initial parameters even for the simulated data.

\subsubsection*{Simulation study: practical identifiability}

We analysed the simulated experiments, aiming to detect modification effects at significance level~$\alpha=0.05$. We assessed practical identifiability based on profile likelihood-based confidence intervals for the model parameters using an adapted threshold $\Delta_{0.05}^{\text{Cant}}=7.16$ instead of $\Delta_{0.05}^{\chi_1^2}=3.84$ in Eq~\eqref{statistics:prof_lik_ci}. Confidence intervals based on this adapted threshold $\text{CI}_{0.95}^{\text{Cant}}(\theta_j)$ accounted for a potential underestimation of parameter uncertainty induced by the asymptotic $\chi_1^2$-distribution assumption in finite-sample cases. We justify this adaptation in a later subsection using data from gene knockout trials. 

The profile likelihood-based confidence intervals $\text{CI}_{0.95}^{\text{Cant}}(\theta_j^{(e)})$, for $j=1,2,3$, revealed that these parameters were practically identifiable for each of the~$1200$ data sets. 
For the estimation of the parameters of the logistic growth model, we set $\theta_4^{(l)}=x_1(0)=5\cdot10^4$ for each mouse-specific version of~$\theta_4^{(l)}$ to ensure structural identifiability. We additionally fixed the carrying capacity and set $\theta_3^{(l)}=K=10^9$ to facilitate parameter estimation. 
The profile likelihood-based confidence intervals showed that the mouse-specific versions of~$\theta_5^{(l)}=x_2(0)$ and~$\theta_6^{(l)}=\sigma$ were practically identifiable for almost all of the~$1200$ data sets. In the absence of a modification effect, proliferation and death rates coincided for both populations, implying $\lambda_1=\lambda_2$. In this case, the growth parameter~$\theta_1^{(l)}$ was structurally non-identifiable and consequently practically non-identifiable (except for five data sets due to numerical artefacts). In contrast, the parameter~$\delta$ was practically identifiable for about three quarters of the data sets across all considered sample sizes. Increasing modification strength and sample size generally improved the practical identifiability of~$\theta_1^{(l)}$. Nevertheless, $\theta_1^{(l)}$ remained practically identifiable in less than half of the data sets across most scenarios and sample sizes (except for the strong modification scenario and $n=32$), whereas~$\delta$ was practically identifiable in substantially more than half of the data sets across nearly all scenarios and sample sizes.

\subsubsection*{Simulation study: exponential and logistic growth models revealed modification effects most reliably}

\begin{figure}[!h]
	\centering
	\includegraphics[width=\textwidth]{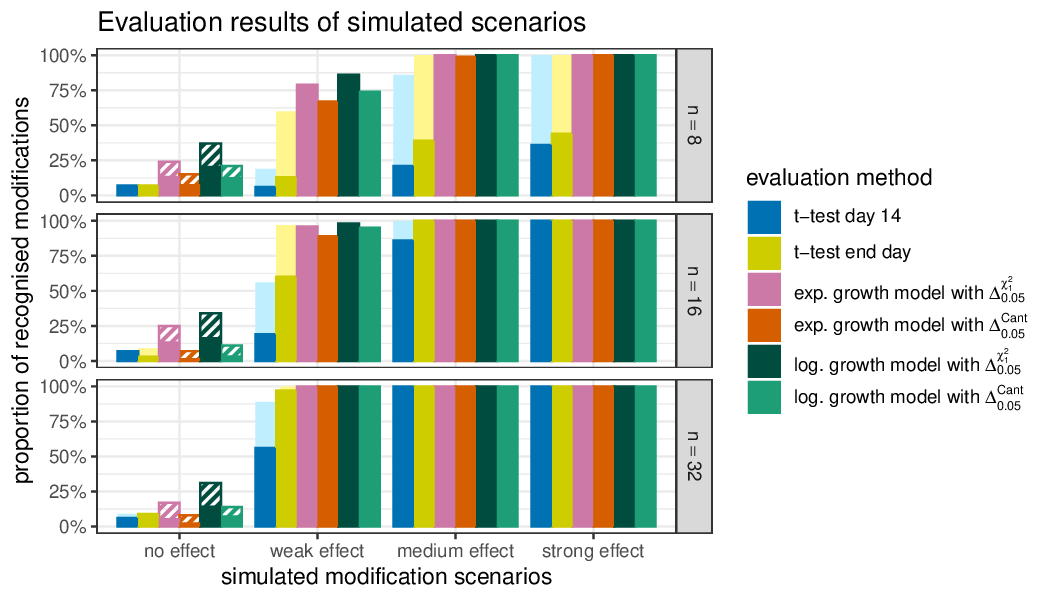}
	\caption{\textbf{Aggregated detection results of the simulation study.}
    Detection results for simulated data via six evaluation methods: paired $t$-tests for two different end days, and the exponential and logistic growth model, for which profile likelihood-based $95$\,\% confidence intervals of~$\theta_1^{(e)}$ and~$\delta$ were constructed using either the quantile $\Delta_{0.05}^{\chi_1^2}=3.84$ or the adapted threshold $\Delta_{0.05}^{\text{Cant}}=7.16$. The data stemmed from four simulated effect size scenarios and three sample size settings ($n\in\{8, 16, 32\}$). Dashed parts represent the proportion of confidence intervals for the corresponding evaluation quantity which are a subset of $\mathbb{R}_{-}$, indicating a growth-enhancing rather than growth-inhibiting effect. Light blue and light yellow bars represent the numbers of detected modification effects of a corresponding scenario for the next larger sample size for paired $t$-tests of day $14$ measurements and endpoint measurements, respectively (considering data sets with sample size~$64$ for $n=32$). Fig~\ref{app:heat_map_simulation} provides a detailed visualisation of detection results per dataset, revealing individual (dis-)agreement between evaluation methods.}
	\label{app:simulation_significant_cases}
\end{figure}

We compared the detection performance of the growth models with paired $t$-tests for simulated experimental scenarios. An appropriate detection method recognised present effects reliably, and if there were no effects in the data-generating process, none should be recognised.
Fig~\ref{app:simulation_significant_cases} displays the proportions of detected effects for all evaluation methods, simulated modification scenarios and sample sizes. An overview of the single detection results is shown in Fig~\ref{app:heat_map_simulation} for each simulated data set and all evaluation methods. 

In the first scenario without any imposed modification (`no effect'), the logistic growth model with quantile~$\Delta_{0.05}^{\chi_1^2}$ showed the highest number of allegedly identified growth-inhibiting effects among the considered evaluation methods and across all sample sizes. Around half of these effects were declared to be growth-enhancing, i.\,e.\ $\text{CI}_{0.95}^{\chi_1^2}(\delta)\subset\mathbb{R}_-$. These proportions are represented by the dashed parts of the bars in Fig~\ref{app:simulation_significant_cases}. For larger sample sizes ($n=16$ and $n=32$), the exponential and logistic growth models with adapted threshold~$\Delta_{0.05}^{\text{Cant}}$ yielded similarly low false-classification rates as the paired $t$-tests and thus became competitive, particularly the exponential growth model. Due to the more conservative construction of the confidence intervals~$\text{CI}_{0.95}^{\text{Cant}}(\theta_1^{(e)})$ and~$\text{CI}_{0.95}^{\text{Cant}}(\delta)$, the set of cases with detected effects based on the adapted threshold~$\Delta_{0.05}^{\text{Cant}}$ was always a subset of the cases classified as growth-inhibiting based on the quantile~$\Delta_{0.05}^{\chi_1^2}$ for both growth models. Furthermore, we found that most of the allegedly identified effects overlapped across the different evaluation methods for the three sample sizes, especially for the growth model-based evaluations (cf.\ Fig~\ref{app:heat_map_simulation}).

If a weak modification effect was present in the data, the logistic growth model with quantile~$\Delta_{0.05}^{\chi_1^2}$ performed best in detecting such effects. When comparing the two growth models for one of the two thresholds, their performance was similar and data sets with detected effects overlapped. For $n=32$, this comparison becomes less informative, as both growth models detected all modifications irrespective of the chosen threshold (cf.\ Fig~\ref{app:simulation_significant_cases}).

Regarding the sample size, it appeared disadvantageous for the $t$-tests to compare their detection performance with the growth model-based approaches from the same indicated sample size, i.\,e.\ comparing the corresponding bars within the same row in Fig~\ref{app:simulation_significant_cases}. That is because of the special design of the simulated datasets, where the output measurements were split into those for day~$14$ and those for day~$40$. The $t$-tests were computed based on only one output measurement time each, whereas the model-based methods took into account measurements of both time points. Thus, the growth models relied on sample sizes twice as large as for the $t$-tests. 
This represents one of the strengths associated with our model-based approaches: they can handle measurements taken at different points in time, as is the case in the motivating dataset of our study.
However, for an alternative comparison, we also included light shaded bars in Fig~\ref{app:simulation_significant_cases} showing the $t$-test results of the next larger sample size. This way, the bars of growth model-based evaluations for a sample size of, e.\,g., $n=8$ can directly be compared with the bars of the $t$-tests for $n=16$.  
This consideration particularly plays a role for the sample size~$n=8$. Accounting for that, the $t$-tests for day~$40$ measurements tended to recognise effects as reliably as the growth models. For larger sample sizes, this consideration becomes negligible.  

In the medium and strong modification scenario, the exponential and logistic growth models performed equally well in recognising almost all modification effects for all sample size settings. For sample size~$n=8$ and the split data sets, both $t$-tests identifed less than half of the modifications of the two scenarios (cf.\ Fig~\ref{app:simulation_significant_cases}). Moreover, the effects identified by the two $t$-tests overlapped only partially (cf.\ Fig~\ref{app:heat_map_simulation}). However, considering larger sample sizes, all considered approaches identified modification effects equally well.

Overall, the logistic growth model with quantile~$\Delta_{0.05}^{\chi_1^2}$ detected existing modification effects in the data the most. The more conservative variant based on the adapted threshold~$\Delta_{0.05}^{\text{Cant}}$ performed only slightly worse. The exponential growth model and paired $t$-tests for day~$40$ measurements showed nearly the same level of detection performance when the actual underlying sample sizes agreed. Paired $t$-tests for day~$14$ measurements showed the weakest detection performance. When no modifications were imposed in the data generation, the exponential growth model with adapted threshold~$\Delta_{0.05}^{\text{Cant}}$ yielded fewer false detections than the logistic counterpart, while the paired $t$-tests produced the fewest.

The data-generating process based on Gillespie's algorithm inherently favoured the logistic growth model over the exponential growth model. Although the logistic growth model achieved the highest detection rates in the simulation study, its parameters were less frequently practically identifiable for the simulated data sets and proved more difficult to estimate than those of the exponential growth model. This particularly affected~$\theta_1^{(l)}=\lambda_1$ and~$\delta$, especially for small sample sizes or in the absence of a modification effect. Even when a modification effect was present and both~$\theta_1^{(l)}$ and~$\delta$ were structurally identifiable, the profile likelihood of~$\theta_1^{(l)}$ was often flat around the maximum likelihood estimate. Extreme parameter estimates were mainly prevented by the $L_2$-constraint added to the likelihood. This issue partly affected~$\delta$. However, in cases of practical non-identifiability, the corresponding confidence interval $\text{CI}_{0.95}^{\text{Cant}}(\delta)$ was typically open towards infinity and thus still excluded zero.

\subsection*{Analysis of gene knockout trials in PDX models}

In preclinical trials, PDX models of acute leukaemia were genetically modified to explore therapeutic targets. We evaluated measurements of a series of gene knockout experiments for five acute lymphocytic leukaemia (ALL) and two acute myeloid leukaemia (AML) PDX models. For each of these samples, up to seven genes were knocked out and the modified cells were injected into mice. The considered seven genes had turned out to be most promising for reducing cell population growth in a CRISPR-Cas9 knockout screening approach with around~$100$ genes. This resulted in a total of~$44$ single knockout experiments. We anonymised the gene names in this work and each gene was assigned a unique identifier, a number between~$1$ and~$7$.

Six of the seven genes were pairwise combined for engraftment for some leukaemia samples such that a mouse was endowed with in total three cell populations, i.\,e.\ two modified populations and the control population, to reduce the number of mice used. The specific gene pairs were gene 1 and gene 4, gene 2 and gene 3, and gene 5 and gene 7 (cf.\ Table~\ref{app:overview_experiments}). This circumstance had no effect on the application of the exponential growth model, as the cell populations behave independently of each other within this model. Within the logistic growth model, however, the different populations compete with each other for the limited capacity in the bone marrow. Following a comparison of the data characteristics, we still followed the two-dimensional logistic growth model rather than employing a model with three state variables~$x_1$, $x_2$, and~$x_3$. This procedure served the goal of showcasing our principle to experimental data. Consequently, we treated all measurements of the~$44$ gene knockout experiments as if they were taken from PDX models endowed with two cell populations.

To perform a specific CRISPR-Cas9-based gene knockout, up to three different sgRNAs were used per knockout experiment. For each experiment, mice were sacrificed on up to two different days in the course of the experiments. One of these days was around day~$14$ after injection of the cells and the second was a later day that was at an advanced stage of leukaemia, which was sample-specific. On these days, cell numbers in the bone marrow of mice were measured by flow cytometry. As the corresponding input measurements were available from day~$0$, up to three measurement times were available per experiment to investigate growth-inhibiting effects. Information about the individual experiments is collected in Table~\ref{app:overview_experiments}.

Table~\ref{app:overview_experiments} also contains $p$-values of paired $t$-tests comparing measurements from an early time point (around day $14$) and from a variable endpoint day with the corresponding input measurements from day~$0$ at a $95$\,\% confidence level. For the variable endpoint measurements, the two-tailed paired $t$-tests detected significant growth-inhibiting effects for the modified cell populations in $40$ of $44$ experiments. For the measurements around day~$14$, the $t$-tests revealed significant growth-inhibiting effects for the cell populations in $33$ of $38$ experiments (six experiments contained less than two measurements for a day around day~$14$, so the paired $t$-test was not applicable in these cases). 

We statistically inferred the parameters of the exponential growth model for all~$44$ knockout data sets. In contrast, we applied the logistic growth model to a subset of only~$38$ knockout experiments with at least~$8$ observations each, which we considered sufficiently large for estimating the more parameter-rich model (cf.\ Table~\ref{app:overview_experiments}). Again, we considered mouse-specific initial parameters for~$\theta_2^{(e)}$ of the exponential and for $\theta_4^{(l)}=x_1(0)$ and $\theta_5^{(l)}=x_2(0)$ of the logistic growth model. 
We fixed the initial condition $\theta_4^{(l)}=x_1(0)$ for each mouse to the absolute input value of the single experiments to make the logistic growth model structurally identifiable for~$\lambda_1\neq\lambda_2$ (cf.\ Table~\ref{app:overview_experiments}). Nevertheless, the property~$\lambda_1=\lambda_2$ might be true for a knockout data set, which would be a data-driven characteristic. However, in general terms, this property has probability zero if probability densities are assigned to the $\mathbb{R}$-valued parameters~$\lambda_1$ and~$\lambda_2$. As long as the parameters are unequal, regardless of how small their difference is, they are structurally identifiable. Moreover, we fixed the carrying capacity~$\theta_3^{(l)}=K$ of the logistic growth model to the maximum cell number that had ever been measured for the single leukaemia samples to further facilitate parameter estimation (cf.\ Table~\ref{app:overview_experiments}).

Before we discuss estimation results and evaluate the knockout experiments, we focus on the asymptotic $\chi_1^2$-distribution assumption from Eq~\eqref{statistics:prof_lik_ci} for profile likelihood-based confidence intervals of the parameters of both growth models.

\subsubsection*{Verification of the approximate $\chi_1^2$-distribution assumption for confidence intervals}

The~$44$ knockout data sets each contained a maximum of~$22$ observations. To ensure valid confidence intervals for these finite-sample cases, we checked whether it was appropriate to assume the asymptotic $\chi_1^2$-distribution assumption for confidence intervals of the parameters of the exponential and logistic growth model  in Eq~\eqref{statistics:prof_lik_ci}. We first estimated~$\bm{\theta}^{(e)}$ for each of the~$44$ knockout data sets and the parameters of the logistic growth model for the considered~$38$ data sets. Within an estimation for a specific knockout experiment, we included all measurements of the different sgRNAs used. In~\cite{Toensing2023}, the authors present a parametric bootstrapping approach to calculate empirical likelihood ratios of the parameters whose cumulative distribution function (ECDF) is compared to the theoretical cumulative distribution function (CDF) of the $\chi_1^2$-distribution. This comparison can be visualised by probability-probability plots (pp-plots), which display the empirical distribution against the theoretical distribution~\cite{Toensing2023}. In an analysis of~$19$ benchmark ODE models from the systems biology literature, they find that the approximate distribution assumption for empirical likelihood ratios of only 52\,\% of the parameters is appropriate. In these cases, the pp-plot graph either lies in the perfect consensus region around the diagonal or in the upper left triangle of the pp-plot. The latter cases are classified as conservative as for those the ECDF is larger than the CDF, which means that profile likelihood-based confidence intervals tend to be too large. However, if the ECDF is considerably smaller for some values than the CDF, the authors distinguish again between two cases in~\cite{Toensing2023}: if the pp-plot graph lies completely in the lower right triangle, the graph is classified as anti-conservative. If the pp-plot graph crosses both regions around the diagonal, the graph is termed alternating. In both cases, confidence intervals might be too small for certain confidence levels. 

The bootstrapping results of the exponential growth model for the experimental data of gene knockouts were even more severe than those from the literature. The pp-plot graphs of all parameters of~$\bm{\theta}^{(e)}$ for the~$44$ data sets were assigned to the anti-conservative class (Fig~\ref{app:fig_exp_model_pp-plots}). If we examined the pp-plot graphs for a 95\,\% confidence level, also all graphs were classified as anti-conservative at this level. In these cases, a confidence interval based on Eq~\eqref{statistics:prof_lik_ci} would have been too small with the quantile~$\Delta_{0.05}^{\chi_1^2}=3.84$. We thus adopted the suggestion from~\cite{Toensing2023} to use an adapted threshold for profile likelihood-based 95\,\% confidence intervals for~$\bm{\theta}^{(e)}$. The threshold~$\Delta_{0.05}^{\text{Cant}}=7.16$ based on Cantelli's inequality yields larger confidence intervals, compared to the asymptotic threshold, accounting for a potential underestimation of parameter uncertainty. We denoted such a confidence interval with adapted threshold by $\text{CI}_{0.95}^{\text{Cant}}(\theta_j^{(e)})$ in the following.

For the logistic growth model, we applied the bootstrapping approach to $\theta_1^{(l)}=\lambda_1$, $\delta$, $\theta_5^{(l)}=x_2(0)$, and $\theta_6^{(l)}=\sigma$, while keeping~$\theta_3^{(l)}=K$ and each mouse-specific~$\theta_4^{(l)}=x_1(0)$ fixed. The resulting pp-plot graphs of the growth parameters~$\lambda_1$ and~$\delta$ were assigned either to the anti-conservative or the alternating class for all~$38$ experiments. Moreover, all~$38$ experiments yielded anti-conservative pp-plot graphs for~$x_2(0)$ and~$\sigma$ (Fig~\ref{app:fig_log_model_pp-plots}). The classifications at a 95\,\% confidence level turned out to be all anti-conservative. According to these results, we also adopted the adapted threshold for 95\,\% confidence intervals of parameters of the logistic growth model. We again denoted such a confidence interval by $\text{CI}_{0.95}^{\text{Cant}}(\theta_j^{(l)})$ and $\text{CI}_{0.95}^{\text{Cant}}(\delta)$.

\subsubsection*{Practical identifiability}

The confidence intervals $\text{CI}_{0.95}^{\text{Cant}}(\theta_j^{(e)})$ of the exponential growth model revealed that~$\theta_1^{(e)}$ was practically identifiable for~$36$ experiments, while the mouse-specific parameters for~$\theta_2^{(e)}$ turned out to be practically identifiable for all considered mice ($350$ in total) and~$\theta_3^{(e)}$ was practically identifiable for all~$44$ experiments. The confidence intervals $\text{CI}_{0.95}^{\text{Cant}}(\theta_j^{(l)})$ for $\theta_5^{(l)}=x_2(0)$ and $\theta_6^{(l)}=\sigma$ provided analogous results: the mouse-specific parameters for~$\theta_5^{(l)}$ were again practically identifiable for all considered mice ($335$ in total) and~$\theta_6^{(l)}$ was practically identifiable for all~$38$ experiments. In contrast, the growth parameters $\theta_1^{(l)}=\lambda_1$ and $\delta$ were less often practically identifiable, namely for only~$12$ and~$14$ experiments, respectively. The relation of the practical identifiability status of the model parameters and the classification of their pp-plot graph was consistent with the results from~\cite{Toensing2023}, whose authors find that the pp-plot graphs of practically non-identifiable parameters are often assigned to the conservative class. Most of the alternating graphs of $\theta_1^{(l)}=\lambda_1$ and~$\delta$ in our case lay mainly in the conservative region (Fig~\ref{app:fig_log_model_pp-plots}). The remaining parameters in our analysis could be practically identified for most of the experiments and their pp-plot graphs were classified as anti-conservative.

\subsubsection*{Detected effects}

For the exponential growth model, we assessed cell population growth-inhibiting effects of gene knockouts based on the confidence interval $\text{CI}_{0.95}^{\text{Cant}}(\theta_1^{(e)})$, as introduced before. The confidence interval excluded zero for~$41$ experiments. In~$40$ of these experiments, the confidence interval was a subset of the positive real line, and we deduced that the specific gene knockout inhibited the growth of cell Population~1 (cf.\ Table~\ref{app:table_KO_significant_cases}). For one experiment (AML-388 6), the interval was a subset of the negative real line, which would have meant that the gene knockout had a growth-boosting effect. This is rather unlikely from a biological perspective and could be due to the fact that the estimation result for this experiment was only based on four measurements. In addition, input and output measurements were of similar magnitude. Applying the asymptotic threshold~$\Delta_{0.05}^{\chi_1^2}=3.84$ for profile likelihood-based confidence intervals would have revealed~$43$ significant knockout effects. These two methods flagged three and one experiments as non-growth-altering, which were among the four non-growth-altering experiments according to two-tailed paired $t$-tests for the variable endpoint measurements. The three knockouts classified as non-growth altering by the exponential growth model using $\Delta_{0.95}^{\text{Cant}}$ had sample sizes of at most $n=6$, and their input and output measurements were again of similar magnitude. Interestingly, from a medical perspective, these knockouts were associated with gene~$6$ (cf.\ Fig~\ref{app:heat_map_ko}). 

\begin{table}[!h]
    \centering
	\caption{\textbf{Aggregated detection results of knockout experiments.}}
	\label{app:table_KO_significant_cases}
	\begin{tabular*}{\textwidth}{@{\extracolsep\fill}lccc}
    \hline
    Detection method & \CellWithForcedBreak{Considered \\ experiments} & \CellWithForcedBreak{Significant \\ experiments} & \CellWithForcedBreak{Of which \\ growth-enhancing} \\ 
    \hline
    $t$-test day 14 & 38 &  33 & - \\ 
    $t$-test end day & 44 &  40 & - \\ 
    exp. growth model with $\Delta_{0.05}^{\chi_1^2}$ & 44 &  43 &   1 \\ 
    exp. growth model with $\Delta_{0.05}^{\text{Cant}}$ &  44 &  41 &   1 \\ 
    log. growth model with $\Delta_{0.05}^{\chi_1^2}$ & 38 &  38 &   0 \\ 
    log. growth model with $\Delta_{0.05}^{\text{Cant}}$ &  38 &  38 &   0 \\
   \hline
    \end{tabular*}
    \begin{flushleft}
        Results for knockout experiments in PDX models: for each detection method, we list the number of considered experiments, the number of recognised significant gene knockout effects, and the number of recognised significant knockout effects that were growth-enhancing, i.\,e.\ for which the confidence interval of the corresponding evaluation method was a subset of~$\mathbb{R}_-$. Fig~\ref{app:heat_map_ko} provides a detailed visualisation of detection results per dataset, revealing individual (dis-)agreement between evaluation methods.
    \end{flushleft}
\end{table}

To evaluate the~$38$ experiments based on the logistic growth model, we again computed confidence intervals $\text{CI}_{0.95}^{\chi_1^2}(\delta)$ and $\text{CI}_{0.95}^{\text{Cant}}(\delta)$. For both versions, the confidence intervals excluded zero for all~$38$ experiments, for which we deduced a significant growth-inhibiting effect based on the logistic growth model (cf.\ Table~\ref{app:table_KO_significant_cases}). All these~$38$ experiments were among the~$41$ experiments which were significantly growth-inhibiting based on the exponential growth model with adapted threshold $\Delta_{0.05}^{\text{Cant}}$. The three experiments classified as non-growth-altering according to $\text{CI}_{0.95}^{\text{Cant}}(\theta_1^{(e)})$ were not analysed with the logistic growth model, as their sample sizes were fewer than eight observations. Consequently, two knockouts that were not inferred by the logistic growth model are classified as growth-inhibiting according to the exponential growth model using $\Delta_{0.95}^{\text{Cant}}$ (cf.\ Fig~\ref{app:heat_map_ko}). 

However, similar technical challenges of estimating the logistic growth model arose for knockout experiments in PDX models, as observed for the simulation study: the profile likelihoods of~$\theta_1^{(l)}$ and~$\delta$ were flat around the maximum likelihood estimate for several of the experiments for which the two parameters were practically non-identifiable. As before, an $L_2$-constraint on the likelihood prevented extreme parameter estimates. For~$\delta$, the profile likelihoods exhibited only a flat trajectory towards infinity, such that the corresponding $\text{CI}_{0.95}^{\text{Cant}}(\delta)$ excluded zero for all experiments.

\begin{figure}[!h]
	\centering
	\includegraphics[width=0.6\textwidth]{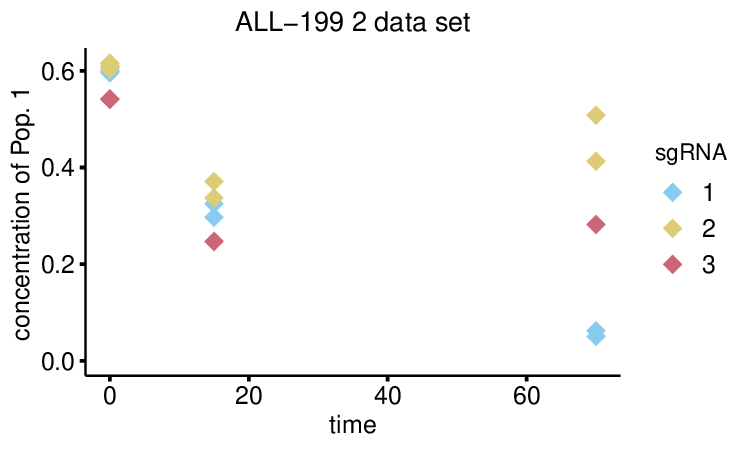}
	\caption{\textbf{Observed concentrations in knockout experiment ALL-199 2.}
    Measurements of the ALL-199 2 experiment, for which three different sgRNAs were used that might have had different knockout efficacy.}
	\label{results:plot_ALL-199_2}
\end{figure}

So far, we have presented results based on combined measurements from up to three different sgRNAs targeting a specific knockout to estimate a single exponential or logistic growth model. The number of used sgRNAs per knockout experiment are collected in Table~\ref{app:overview_experiments}. Even though the most suitable sgRNAs were selected to knock out a gene, the chosen sgRNAs might have differed in their knockout efficacy, as it could have been the case for the ALL-199 2 experiment (Fig~\ref{results:plot_ALL-199_2}). To evaluate the knockouts for each considered sgRNA separately, we re-estimated the exponential growth model for each sgRNA used for a knockout for which more than one sgRNA were used in total. Again, we excluded the experiments where an sgRNA was only applied to one mouse and thus only two observations were obtained. Hence, we considered a total of~$113$ data sets for~$38$ knockouts, with sample sizes ranging from four to eight measurements. Parameter~$\theta_1^{(e)}$ was practically identifiable for~$82$ of these data sets. 
For~$99$ of the estimated models, we found $\text{CI}_{0.95}^{\text{Cant}}(\theta_1^{(e)})\subset\mathbb{R}_+$ and we deduced significant cell population growth-inhibiting effects for these knockouts and the single sgRNAs. Comparing the single evaluations for all sgRNAs used with respect to a specific knockout, we found that the significance of sgRNA sub-effects varied for~$10$ of the considered~$38$ knockouts: for these, the confidence interval $\text{CI}_{0.95}^{\text{Cant}}(\theta_1^{(e)})$ was a subset of the positive real line for at least one sgRNA and for at least one other sgRNA, the interval overlapped zero. As far as hypothesis tests were concerned, we could apply a paired $t$-test for the endpoint measurements to only~$62$ of the~$113$ data sets, as these data sets contained endpoint measurements of more than one mouse. In~$46$ of these cases, the $t$-tests for endpoint measurements provided a significant effect of gene knockout. 
As most of the separated estimations were based on data sets with fewer than eight observations, we refrained from conducting a separate sgRNA analysis with the logistic growth model, as the small sample sizes rendered parameter estimation computationally challenging.

\section*{Discussion and conclusion}

We investigated whether incorporating mechanistic knowledge makes the detection of differences between underlying processes more reliable than a statistical test. We related our analysis to gene knockouts in PDX samples of acute leukaemia as part of preclinical research. To describe the growth of leukaemia cell populations and to detect cell growth-inhibiting effects, we applied an exponential and a logistic growth model. We compared their detection performance with paired $t$-tests, which are often applied as standard, in several ways. 

When applied to real data, we could observe that different detection methods led to different results. However, in the absence of ground truth, this comparison did not yet allow us to determine which method was more reliable. In addition, the performance directly depended on sample sizes. For that reason, we conducted a simulation study to compare the growth models with $t$-tests for analysing time-resolved data of a stochastic process related to gene modification experiments. Our results indicated that the growth models performed well in evaluating the considered type of data sets. When modifications were present, the logistic model achieved higher detection rates, whereas the exponential growth model yielded fewer false-positive classifications in a scenario without imposed modification.

Although the logistic growth model detected numerous effects and approximately reflected the data generating process, its parameters were more challenging to estimate for the considered data type. The profile likelihoods of the growth parameters -- particularly~$\theta_1^{(l)}=\lambda_1$ -- were often flat around the estimates, and only the $L_2$-constraint on the likelihood prevented extreme parameter estimates. This should be taken into account when interpreting both the evaluation results and, in particular, specific parameter estimates.

In the simulation study, the paired $t$-tests for the variable endpoint measurements performed comparably to the growth models, provided that the sample size was sufficiently large: in that case, the $t$-test recognised approximately as many effects as the growth models with adapted threshold~$\Delta_{0.05}^{\text{Cant}}$. The $t$-test benefited from the allocation of the output measurements to a maximum of two measurement days. Thus, a paired $t$-test for late measurement times is an alternative to model-based approaches, both in the presence and absence of growth-inhibiting effects. However, the possibility of combining measurements into a single sample for the $t$-tests depends on the experimental conditions, which typically cannot be influenced at the time of data analysis. This represents a strength of the model-based methods.

For the knockout experiments in PDX models, the key comparison is between the exponential and logistic growth models with adapted threshold~$\Delta_{0.05}^{\text{Cant}}$, and the paired $t$-test for the endpoint measurements: the first method was effective at not flagging non-existing effects in the simulation study, the second reliably detected existing ones, and the third represented the established approach in practice. The framework of the knockout experiments aligned with the first two sample size settings of the simulation study ($n\in\{8,16\}$), as the knockout experiments included at most $n=22$ measurements. The conservative confidence intervals~$\text{CI}_{0.95}^{\text{Cant}}(\theta_1^{(e)})$ and~$\text{CI}_{0.95}^{\text{Cant}}(\delta)$ for the knockout experiments suggested that the range of assumed effect sizes considered in the simulation study (from ``none" to ``strong") constituted a realistic representation of the real-data experiments, although effects exceeding the ``strong" category were observed. However, the true effects in the knockout data sets remained unknown. The exponential growth model and the paired $t$-test were applied to all~$44$ experiments, whereas the logistic growth was applied to a subset of~$38$ experiments. The exponential growth model detected~$41$ effects (of which one was growth-enhancing) and the logistic growth model detected effects in all considered~$38$ data sets, while the paired $t$-test detected~$40$ effects. 
The three knockouts classified as non-growth-altering by the exponential growth model with adapted threshold~$\Delta_{0.05}^{\text{Cant}}$ were likewise identified as non-growth-altering by the paired $t$-test (cf.\ Fig~\ref{app:heat_map_ko}). 
Overall, the effect detection results for the knockout data sets were largely consistent across the three methods.

Another strength of our approach lies in explicitly accounting for the paired structure of the data, as each input–output measurement pair originated from the same mouse. While assuming that a given gene knockout exerted a consistent growth-inhibiting effect across mice, we incorporated mouse-specific heterogeneity by estimating individual initial parameters. In doing so, we could account for differences in the number of cells that ultimately home to the bone marrow~\cite{Ebinger2016,Bahrami2023}.

We further found for the experiments in PDX models that a separate consideration of the sgRNAs used for a knockout seems reasonable, as the sgRNAs might have differed in their knockout efficacy according to the exponential growth model. The confidence interval $\text{CI}_{0.95}^{\text{Cant}}(\theta_1^{(e)})$ could be calculated for these small sample size settings while reflecting parameter uncertainty more appropriately. In contrast, the sample sizes were too small for the more parameter-rich logistic growth model to produce reliable estimation results.

Another limitation of the logistic growth model, particularly in the context of PDX experiments, is that parameter values needed to be fixed. By setting the initial condition $\theta_4^{(l)}=x_1(0)$ equal to the absolute input value of a knockout experiment, only the second initial condition $\theta_5^{(l)}=x_2(0)$ remained to be estimated, thereby reducing the model’s ability to capture mouse-specific heterogeneity. Moreover, the maximum cell numbers were rough guesses for the carrying capacity~$\theta_3^{(l)}=K$. These determinations further affected the estimation results. Based on our preliminary estimations, we expect the same technical difficulties with an analogue two-dimensional Gompertz growth model~\cite{Laird1964a} due to its similarity to the logistic growth model. Nevertheless, parameter estimation for these growth models might be improved, if, for example, the selection of measurement time points is optimised or absolute cell numbers are available. The same might apply to other dynamic growth models that have been used in the past to model cancer growth~\cite{Benzekry2014}.

In~\cite{Kreutz2007}, a multiplicative log-normally distributed measurement error is shown to adequately capture biological variability and experimental noise for measurements which are non-negative concentrations. However, in our case, the variability observed in the knockout measurements did not follow a multiplicative log-normal pattern but was more consistent with an additive normal distribution. Therefore, we employed an additive, normally distributed error model on the modelled observable throughout this article (cf.\ Eq~\eqref{statistics:observable_error}). 

In conclusion, we showed that dynamic models provide a more informative approach to analyse time-resolved data than a non-dynamic statistical test. In our application, both the exponential and the logistic growth models proved suitable for evaluating the considered gene modification experiments. The exponential growth model yielded more stable results, particularly in the presence of weak or absent effects and in small-sample settings. Thus, both growth models can serve as a basis to optimise the experimental design of such experiments, enabling the use of as few mice as necessary while allocating measurement times efficiently. 

\nolinenumbers

\begin{appendices}

\setcounter{figure}{6} 
\renewcommand{\thefigure}{\arabic{figure}}

\setcounter{table}{2}
\renewcommand{\thetable}{\arabic{table}}

\section*{Supporting information}

\begin{table}[!h]
\begin{adjustwidth}{-2.25in}{0in}
    \centering
	\caption{\textbf{Details about gene knockout experiments.}}
	\begin{tabular*}{\textwidth}{@{\extracolsep\fill}lccccccc}
		\toprule
        \multicolumn{6}{@{}c@{}}{Details of experiment} & \multicolumn{2}{@{}c@{}}{$p$-value of $t$-tests}
        \\\cmidrule{1-6} \cmidrule{7-8}
		Sample & Gene & \#sgRNAs & 
        \CellWithForcedBreak{Sample \\ size} &
        \CellWithForcedBreak{Max. cell \\ number} &
        \CellWithForcedBreak{Input \\ value} &
        day 14 & end day \\ 
		\midrule
        ALL-1034 &    1 &    3 &   18 &  110 &   33 & 0.001 & 0.006 \\ 
  ALL-1034 &    2 &    3 &   18 &  110 &   33 & 0.008 & 0.001 \\ 
  ALL-1034 &    3 &    3 &   18 &  110 &   33 & 0.000 & 0.000 \\ 
  ALL-1034 &    4 &    3 &   18 &  110 &   33 & 0.006 & 0.004 \\ 
  ALL-1034 &    5 &    3 &   20 &  110 &   33 & 0.000 & 0.000 \\ 
  ALL-1034 &    6 &    1 &    4 &  110 &   33 & - & 0.747 \\ 
  ALL-1034 &    7 &    3 &   20 &  110 &   33 & 0.000 & 0.000 \\ 
  ALL-199 &    1 &    3 &   20 &   70 &  400 & 0.000 & 0.000 \\ 
  ALL-199 &    2 &    3 &   20 &   70 &  400 & 0.000 & 0.023 \\ 
  ALL-199 &    3 &    3 &   20 &   70 &  400 & 0.033 & 0.000 \\ 
  ALL-199 &    4 &    3 &   20 &   70 &  400 & 0.017 & 0.012 \\ 
  ALL-199 &    5 &    3 &   20 &   70 &  400 & 0.000 & 0.000 \\ 
  ALL-199 &    7 &    3 &   20 &   70 &  400 & 0.000 & 0.000 \\ 
  ALL-265 &    1 &    3 &   20 &   70 &  100 & 0.000 & 0.000 \\ 
  ALL-265 &    2 &    3 &   16 &   70 &  100 & 0.001 & 0.000 \\ 
  ALL-265 &    3 &    3 &   16 &   70 &  100 & 0.000 & 0.000 \\ 
  ALL-265 &    5 &    2 &    8 &   70 &  100 & 0.004 & 0.013 \\ 
  ALL-265 &    7 &    3 &   12 &   70 &  100 & 0.007 & 0.023 \\ 
  ALL-50 &    1 &    3 &   14 &  260 &   30 & 0.000 & 0.000 \\ 
  ALL-50 &    2 &    3 &   14 &  260 &   30 & 0.015 & 0.000 \\ 
  ALL-50 &    3 &    3 &   14 &  260 &   30 & 0.000 & 0.000 \\ 
  ALL-50 &    4 &    3 &   12 &  260 &   30 & 0.053 & 0.025 \\ 
  ALL-50 &    5 &    3 &   18 &  260 &   30 & 0.001 & 0.000 \\ 
  ALL-50 &    6 &    1 &    6 &  260 &   30 & - & 0.168 \\ 
  ALL-50 &    7 &    3 &   18 &  260 &   30 & 0.003 & 0.000 \\ 
  ALL-502 &    1 &    3 &   16 &  100 &   40 & 0.219 & 0.004 \\ 
  ALL-502 &    2 &    3 &   12 &  100 &   40 & 0.073 & 0.026 \\ 
  ALL-502 &    3 &    3 &   20 &  100 &   40 & 0.012 & 0.002 \\ 
  ALL-502 &    4 &    3 &   16 &  100 &   40 & 0.013 & 0.009 \\ 
  ALL-502 &    5 &    3 &   18 &  100 &   40 & 0.003 & 0.016 \\ 
  ALL-502 &    6 &    1 &    4 &  100 &   40 & - & 0.407 \\ 
  ALL-502 &    7 &    3 &   20 &  100 &   40 & 0.000 & 0.000 \\ 
  AML-356 &    1 &    3 &   22 &   70 &  400 & 0.001 & 0.000 \\ 
  AML-356 &    2 &    3 &    6 &   70 &  400 & - & 0.018 \\ 
  AML-356 &    3 &    3 &    6 &   70 &  400 & - & 0.013 \\ 
  AML-356 &    5 &    3 &   20 &   70 &  400 & 0.000 & 0.000 \\ 
  AML-356 &    7 &    3 &   20 &   70 &  400 & 0.000 & 0.000 \\ 
  AML-388 &    1 &    3 &   16 &   80 &  400 & 0.099 & 0.018 \\ 
  AML-388 &    2 &    3 &   20 &   80 &  400 & 0.013 & 0.006 \\ 
  AML-388 &    3 &    3 &   20 &   80 &  400 & 0.000 & 0.000 \\ 
  AML-388 &    4 &    3 &   16 &   80 &  400 & 0.105 & 0.161 \\ 
  AML-388 &    5 &    3 &   20 &   80 &  400 & 0.010 & 0.015 \\ 
  AML-388 &    6 &    1 &    4 &   80 &  400 & - & 0.009 \\ 
  AML-388 &    7 &    3 &   20 &   80 &  400 & 0.000 & 0.001 \\  
        \hline
	\end{tabular*}
\begin{flushleft}
Overview of all gene knockout experiments that were analysed. From left to right: leukaemia sample name, identifier of knocked-out gene, number of sgRNAs used, sample size, maximum number of cells in the sample that had ever been measured in the bone marrow (BM) of a PDX model in the laboratory where the experiments were conducted  (in million), absolute number of input cells (in thousands), and $p$-value for the two $t$-tests comparing input measurements from day zero with corresponding measurements from (approx.) day $14$ and a variable end day.
\end{flushleft}
\label{app:overview_experiments}
\end{adjustwidth}
\end{table}

\clearpage

\begin{figure}[h]
	\centering
	\includegraphics[width=0.8\textwidth]{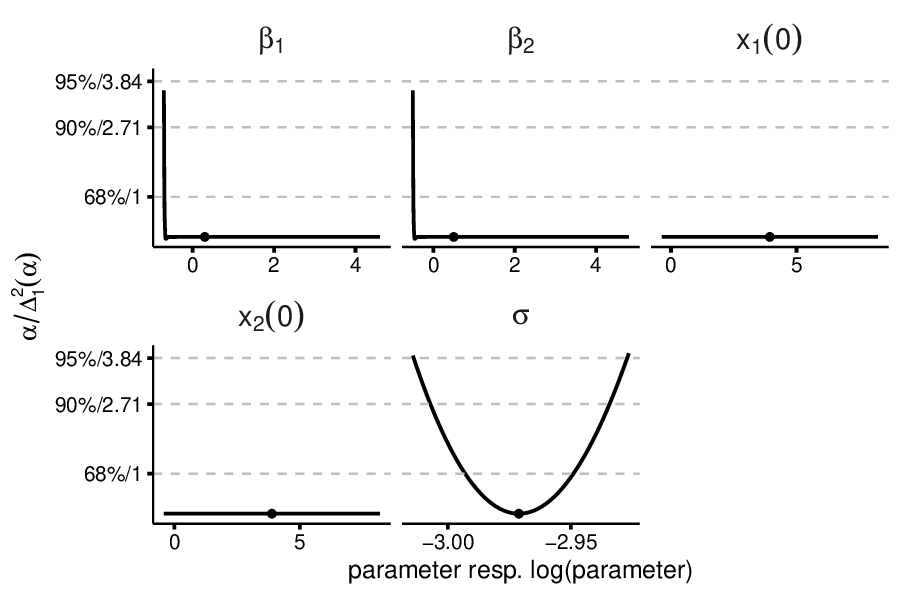}
	\caption{\textbf{Profile likelihoods of non-identifiable exponential growth model.}
    Profile likelihoods for $\bm{\theta}^{(e)}=(\beta_1,\beta_2,x_1(0),x_2(0),\sigma)$ of the exponential growth model, where $x_1(0)$, $x_2(0)$ and $\sigma$ were log-transformed, for simulated data with $\bm{\theta}^{(e)}=(0.3,0.5,50,50,0.05)$, showing that all parameters were structurally non-identifiable except $\sigma$.}
	\label{app:fig_exp_model_non-ident_model}
\end{figure}

\begin{figure}[h]
	\centering
	\includegraphics[width=0.8\textwidth]{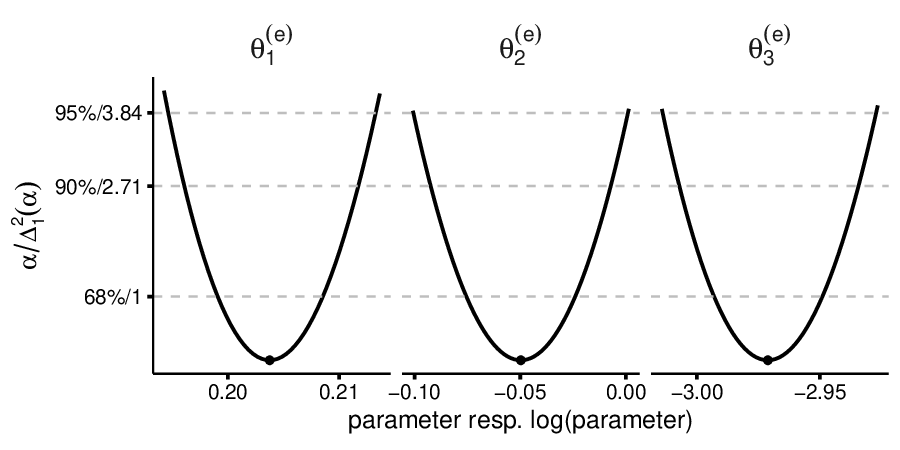}
	\caption{\textbf{Profile likelihoods of identifiable exponential growth model.}
    Profile likelihoods for $\bm{\theta}^{(e)}=(\theta_1^{(e)},\theta_2^{(e)},\theta_3^{(e)})$ of the reparametrised exponential growth model, where $\theta_2^{(e)}$ and $\theta_3^{(e)}$ were log-transformed, based on the same data set as for Fig~\ref{app:fig_exp_model_non-ident_model}, showing a structurally identifiable model parametrisation.}
	\label{app:fig_exp_model_ident_model}
\end{figure}

\begin{figure}[h]
	\centering
	\includegraphics[width=0.8\textwidth]{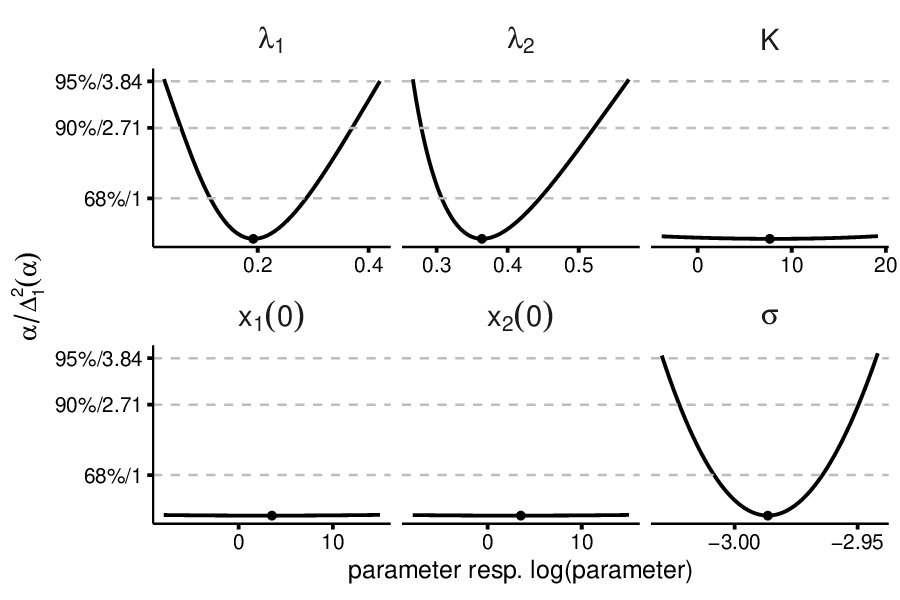}
	\caption{\textbf{Profile likelihoods of non-identifiable logistic growth model.}
    Profile likelihoods for $\bm{\theta}^{(l)}=(\lambda_1,\lambda_2,K,x_1(0),x_2(0),\sigma)$ of the logistic growth model, where $K$, $x_1(0)$, $x_2(0)$ and $\sigma$ were log-transformed, for simulated data with $\bm{\theta}^{(l)}=(0.1,0.3,1000,50,50,0.05)$, showing that $K$, $x_1(0)$ and $x_2(0)$ were structurally non-identifiable.}
	\label{app:fig_log_model_non-ident_model}
\end{figure}

\begin{figure}[h]
	\centering
	\includegraphics[width=0.8\textwidth]{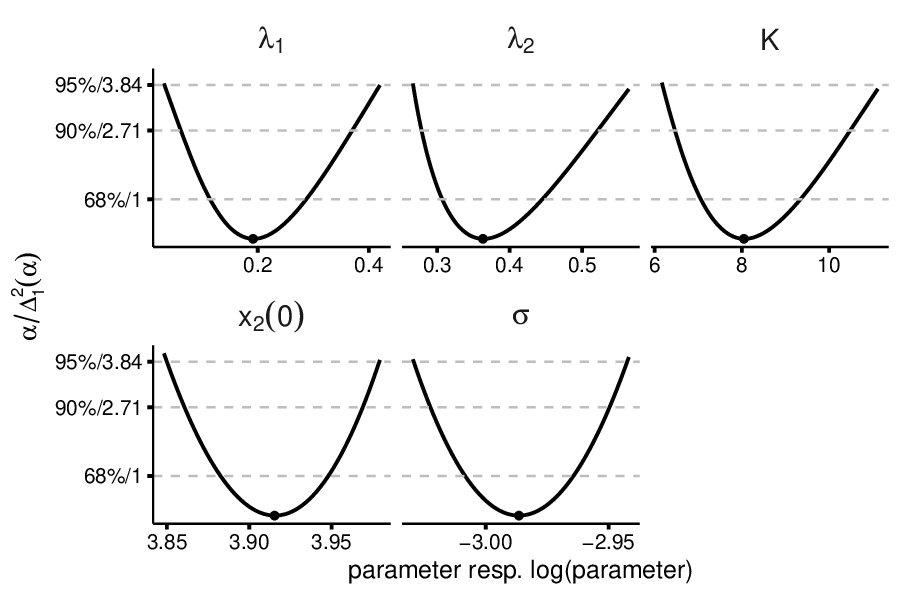}
	\caption{\textbf{Profile likelihoods of identifiable logistic growth model.}
    Profile likelihoods for $\bm{\theta}^{(l)}$ of the logistic growth model, where $K$, $x_2(0)$ and~$\sigma$ were $\log$-transformed and $x_1(0)=50$ was fixed, based on the same data set as for Fig~\ref{app:fig_log_model_non-ident_model}, showing a structurally identifiable model parametrisation.}
	\label{app:fig_log_model_ident_model}
\end{figure}

\begin{figure}[h]
	\centering
	\includegraphics[width=0.8\textwidth]{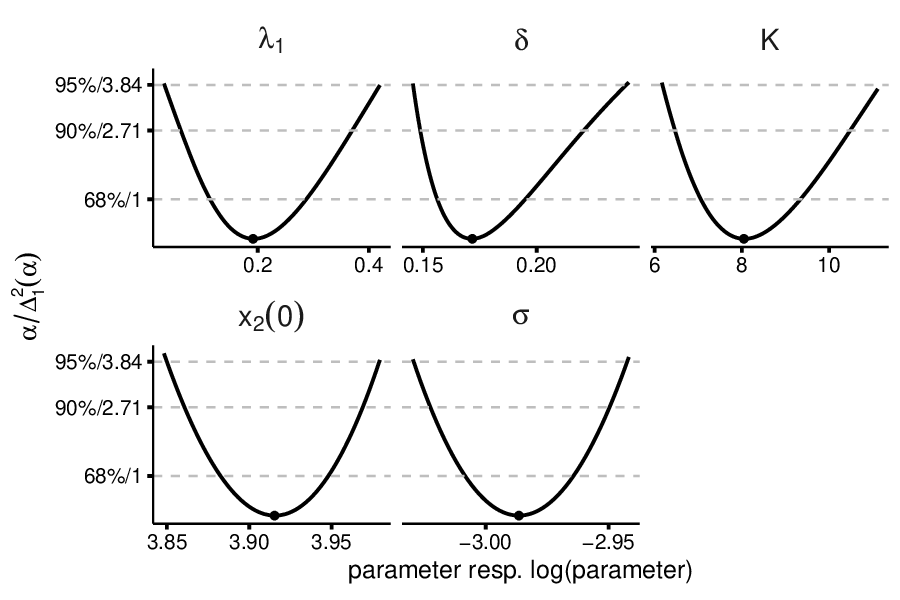}
	\caption{\textbf{Profile likelihoods of identifiable logistic growth model with $\delta=\lambda_2-\lambda_1$.}
    Profile likelihoods for the parameters of the logistic growth model with~$\lambda_2$ substituted by $\lambda_2=\delta+\lambda_1$, where $K$, $x_2(0)$ and~$\sigma$ were $\log$-transformed and $x_1(0)=50$ was fixed, based on the same data set as for Fig~\ref{app:fig_log_model_non-ident_model}, showing a structurally identifiable model parametrisation.}
	\label{app:fig_log_model_ident_model_delta}
\end{figure}

\begin{figure}[h]
	\centering 
	\includegraphics[width=\textwidth]{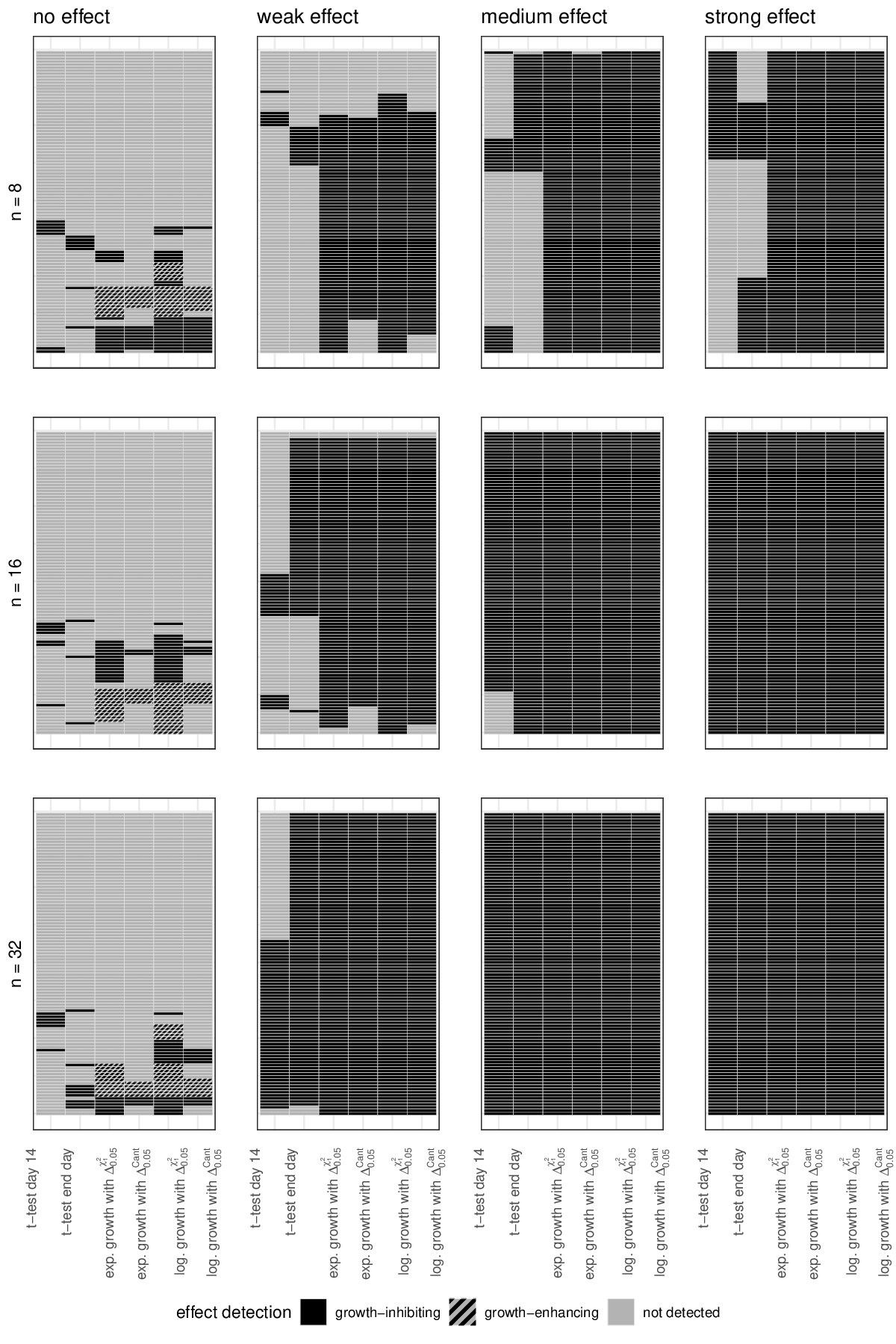}
	\caption{\textbf{Detection results for all simulated data sets.}
    More detailed presentation of detection results from Fig~\ref{app:simulation_significant_cases}: while the other figure displays accumulated numbers of detected effects per evaluation method and sample size, this plot reveals the (dis-)agreement between evaluation methods for each of the 100 datasets. The evaluation results are displayed in the rows of the corresponding subplot. Recognised growth-altering effects in data sets are marked in black (dashed for growth-enhancing). If no effect was detected, the data set is marked in grey. Rows are sorted so that data sets with similar evaluation results are grouped together.}
	\label{app:heat_map_simulation}
\end{figure}

\begin{figure}[h]
	\centering 
	\includegraphics[width=0.8\textwidth]{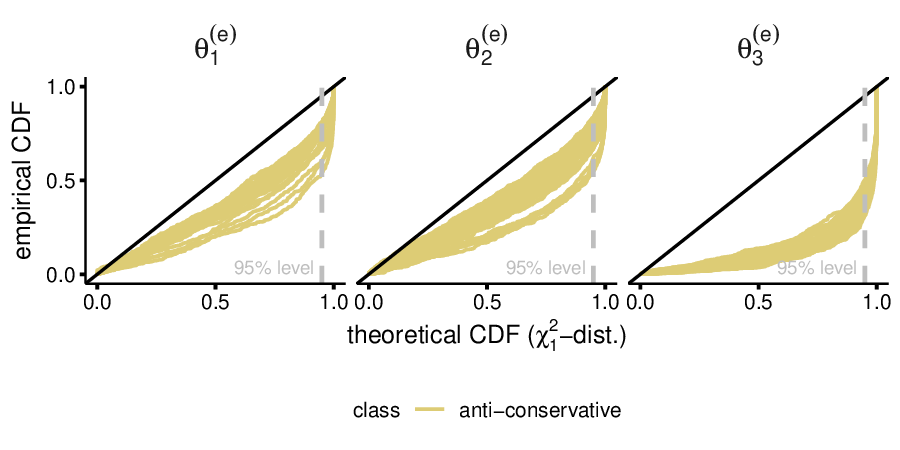}
	\caption{\textbf{Probability-probability plots for the exponential growth model.}
    Probability-probability plots of the components of the parameter vector~$\bm{\theta}^{(e)}$ of the exponential growth model for all $44$ knockout experiments in PDX models, showing that all pp-plots were assigned to the anti-conservative class, even at a confidence level of $0.95$. Note that the pp-plot graph for each mouse specific initial parameter is included in the plot for~$\theta_2^{(e)}$ (350 in total).}
	\label{app:fig_exp_model_pp-plots}
\end{figure}

\begin{figure}[h]
	\centering 
	\includegraphics[width=\textwidth]{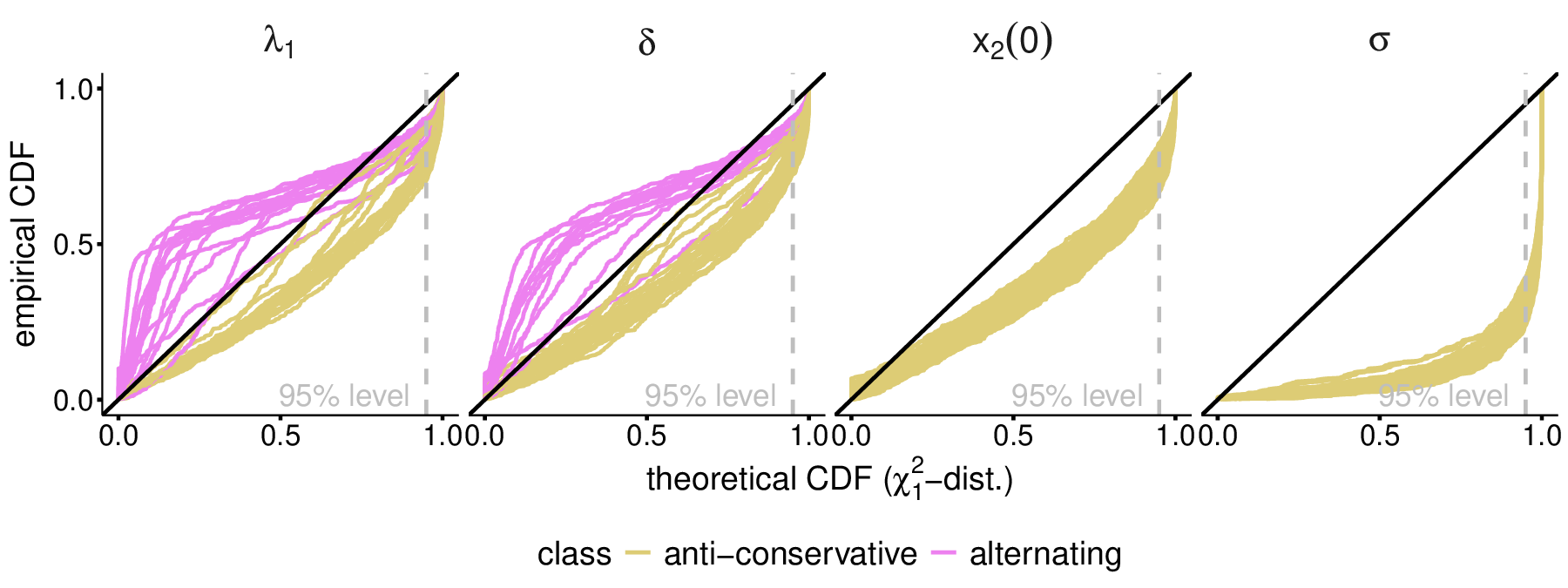}
	\caption{\textbf{Probability-probability plots for the logistic growth model.}
    Probability-probability plots of the parameters~$\theta_1^{(l)}=\lambda_1$, $\delta$, $\theta_5^{(l)}=x_2(0)$, and~$\theta_6^{(l)}=\sigma$ of the logistic growth model for~$38$ knockout experiments in PDX models while keeping $\theta_3^{(l)}=K$ and $\theta_4^{(l)}=x_1(0)$ fixed. Most pp-plots of $\lambda_1$ and $\delta$ were assigned to the anti-conservative class, while all pp-plots of $x_2(0)$ and $\sigma$ were assigned to the anti-conservative class. An anti-conservative classification for each parameter was present at a confidence level of $0.95$. Note that the pp-plot graph for each mouse specific initial parameter is included in the plot for~$x_2(0)$ (335 in total).}
	\label{app:fig_log_model_pp-plots}
\end{figure}

\begin{figure}[h]
	\centering 
	\includegraphics[width=\textwidth]{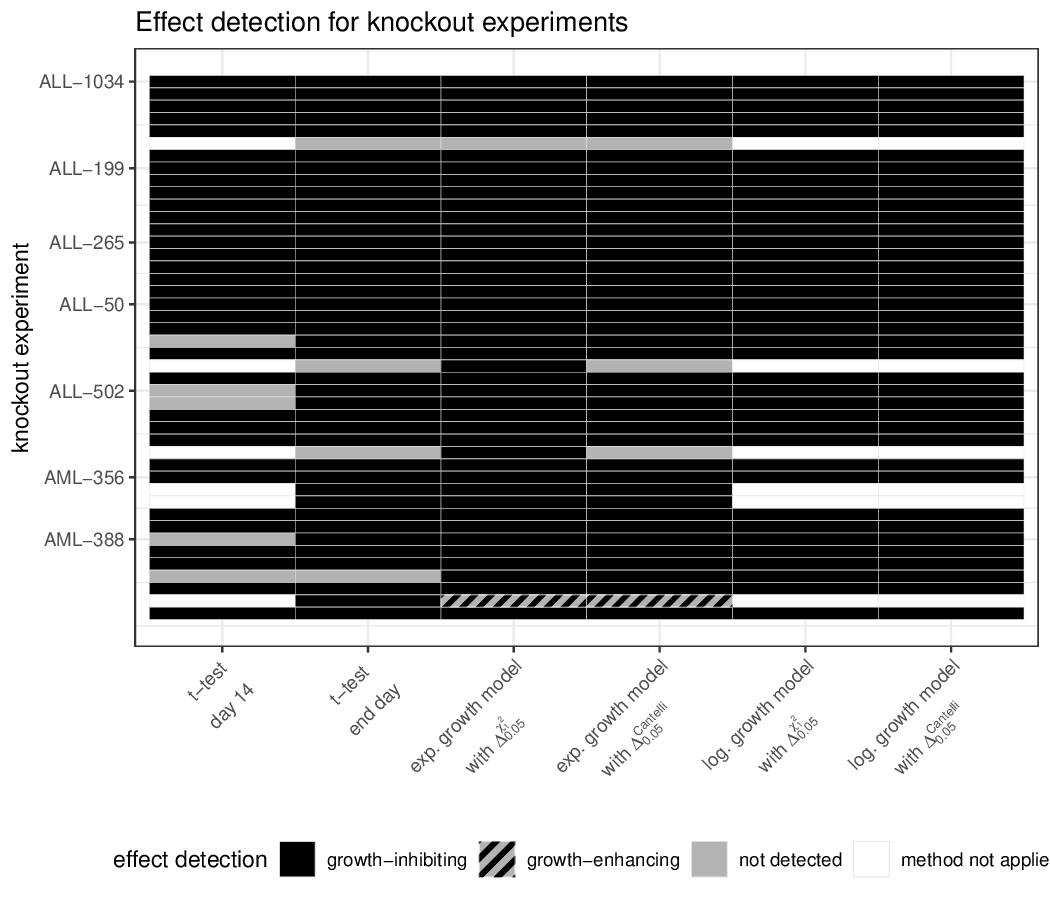}
	\caption{\textbf{Detection results for the knockout experiments.}
    More detailed presentation of detection results from Table~\ref{app:table_KO_significant_cases}: while the table displays accumulated numbers of detected effects for knockout experiments in PDX models via six evaluation methods, this plot reveals the (dis-)agreement between evaluation methods for each dataset. The experiments are arranged in rows as shown in Table~\ref{app:overview_experiments}, i.\,e.\ in ascending order according to the gene numbers for each leukaemia sample considered. Experiments for which significant growth-altering effects were recognised based on an evaluation method are marked in black (black-striped if growth-enhancing). If no effect was detected, the experiment is marked in grey. If the sample size of an experiment was too small to apply a method, it is marked in white.}
	\label{app:heat_map_ko}
\end{figure}

\end{appendices}

\clearpage 

\paragraph{Acknowledgements} 
We thank Thanh Thao Bui for valuable assistance in implementing the models. The authors used ChatGPT (OpenAI, GPT-5.2) for linguistic refinement. The authors reviewed and revised all AI-generated suggestions and take full responsibility for the final content.

\paragraph{Author contributions}
Conceptualisation: Julian W\"asche, Christiane Fuchs; Formal analysis: Julian W\"asche; Funding acquisition: Christiane Fuchs, Irmela Jeremias; Investigation: Romina Ludwig, Irmela Jeremias; Methodology: Julian W\"asche, Christiane Fuchs; Supervision: Christiane Fuchs; Visualisation: Julian W\"asche; Writing - original draft preparation: Julian W\"asche; Writing - review and editing: Julian W\"asche, Christiane Fuchs, Romina Ludwig, Irmela Jeremias.

\paragraph{Data availability}
All simulations and all statistical analyses were performed in~\textsf{R} (Version 4.5.0). Code for the statistical analyses, analysis results and the data sets are available at \url{https://github.com/fuchslab/Modelling_Evaluation_Leukaemia_Trials}.

\paragraph{Funding}
The project ``Opti-Trials'' underlying this publication was funded by the German Federal Ministry of Research, Technology and Space (BMFTR) under grant number 16DKWN112A. The responsibility for the content of this publication lies solely with the author(s). Co‑funded by the European Union -- NextGenerationEU.

\paragraph{Conflicts of interest}
The authors declare no conflicts of interest.

\bibliography{references}

\end{document}